\documentclass[sigconf]{acmart}
\usepackage{amsmath,amsfonts}
\usepackage{adjustbox}

\usepackage{balance}
\usepackage{algorithm}
\usepackage[noend]{algpseudocode}
\usepackage{graphicx}
\usepackage{textcomp}
\usepackage{xcolor}
\usepackage{caption}
\usepackage{subcaption}
\usepackage{tabularx}
\usepackage{multirow}
\usepackage{makecell}
\usepackage{diagbox}
\usepackage{booktabs}
\usepackage{color, colortbl}
\usepackage{courier}
\usepackage{pifont}
\definecolor{Gray}{gray}{0.9}
\algrenewcommand\algorithmicrequire{\textbf{Input:}}
\algrenewcommand\algorithmicensure{\textbf{Output:}}

\AtBeginDocument{%
 }

\author{Yoonhyuk Choi}
\orcid{0000-0003-4359-5596}
\affiliation{
  \institution{
  Samsung}
  \city{Seoul}
  \country{Republic of Korea}
}
\email{chldbsgur123@gmail.com}

\author{Jiho Choi}
\orcid{0000-0002-7140-7962}
\affiliation{
  \institution{
  KAIST}
  \city{Seoul}
  \country{Republic of Korea}
}
\email{jihochoi @ kaist.ac.kr}

\author{Taewook Ko}
\orcid{0000-0001-7248-4751}
\affiliation{
  \institution{
  Samsung}
  \city{Seoul}
  \country{Republic of Korea}
}
\email{taewook.ko @ snu.ac.kr}

\author{Chong-Kwon Kim}
\orcid{0000-0002-9492-6546}
\affiliation{
  \institution{Korea Institute of Energy Technology}
  \city{Naju}
  \country{Republic of Korea}
}
\email{ckim @ kentech.ac.kr}

\renewcommand{\shortauthors}{Yoonhyuk Choi, Jiho Choi, Taewook Ko, and Chong-Kwon Kim}

\newtheorem{manualtheoreminner}{\textbf{Proof of proposition}}
\newenvironment{manualtheorem}[1]{%
  \renewcommand\themanualtheoreminner{#1}%
  \manualtheoreminner
}{\endmanualtheoreminner}

\newtheorem{manualproposition}{\textbf{Proposition}}
\newenvironment{manualprop}[1]{%
  \renewcommand\themanualproposition{#1}%
  \manualproposition
}{\endmanualproposition}

\begin{document}

\title{Review-Based Hyperbolic Cross-Domain Recommendation}

\begin{abstract} 
The issue of data sparsity poses a significant challenge to recommender systems. Recently, algorithms that leverage side information (review texts) or Cross-Domain Recommendation (CDR) have emerged. Nevertheless, existing methodologies assume an Euclidean embedding space, encountering difficulties in accurately representing richer text information and managing complex user-item interactions. This paper advocates a hyperbolic CDR approach for modeling review-based user-item relationships. We first emphasize that conventional distance-based domain alignment techniques may cause problems because small modifications in hyperbolic geometry result in magnified perturbations, ultimately leading to the collapse of hierarchical structures. To address this challenge, we propose hierarchy-aware embedding and domain alignment schemes that adjust the scale to extract domain-shareable information without disrupting structural forms. Extensive experiments substantiate the efficiency, robustness, and scalability of the proposed model. The source code is given \textit{here}\footnote{https://github.com/ChoiYoonHyuk/HEAD}.
\end{abstract}

\ccsdesc[500]{Information systems~Recommender systems}

\keywords{Recommender system, review-based cross-domain recommendation, hyperbolic embedding, hierarchy-aware domain alignment}

\maketitle

\begin{figure}[t]
  \centering
  \vspace{2mm}
  \includegraphics[width=0.49\textwidth]{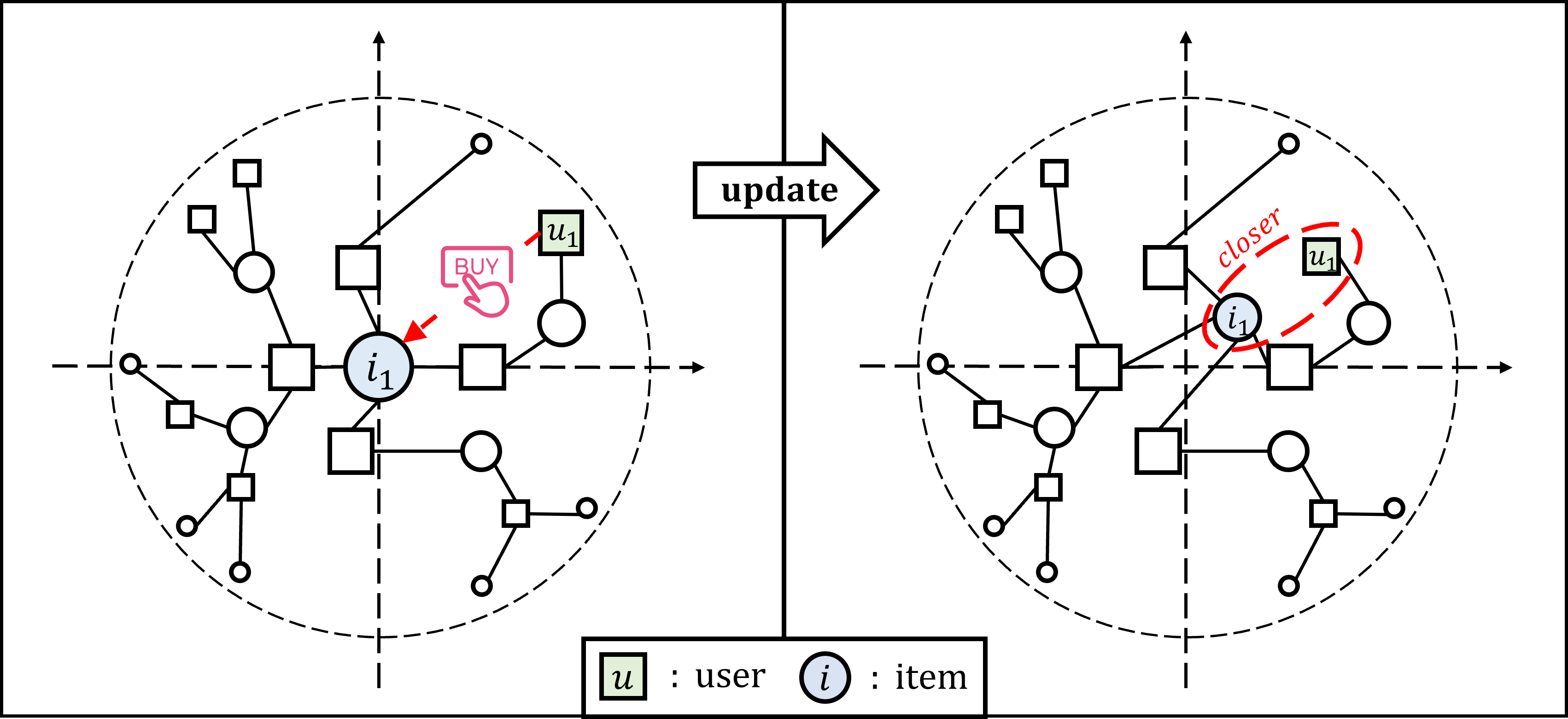}
  \caption{The geometric properties of hyperbolic space require that popular (or most interacted) nodes be placed near the origin. Let us assume that user $u_1$ purchased an item $i_1$ (left). Given that general algorithms bring relevant nodes closer together, after the update, a structural collapse occurs (right) as $i_1$ moves farther from the origin}
  \label{euc_hyp}
\end{figure}

\section{Introduction}
A recommender system has evolved into a fundamental tool across real-world applications \cite{lu2015recommender,zhang2019deep} such as Amazon, and Tripadvisor. Despite remarkable popularity and commercial successes, the performance can be inflicted heavily in data-scarce scenarios. The impediments of data sparsity encompass issues like cold-start problems, which have gained academic focus recently. Numerous research endeavors have employed various forms of side information ranging from social relationships \cite{kazienko2011multidimensional}, and hierarchical interactions \cite{liu2019recommender} to item images \cite{addagarla2020probabilistic}. In particular, textual data in the form of reviews has become one of the most extensively utilized sources \cite{zheng2017joint,al2019multi,srifi2020recommender,choi2022based}. These endeavors have demonstrated a measure of effectiveness in ameliorating sparsity concerns. However, inherent limitations persist in addressing fundamental issues, especially when the extent of interaction is insufficient.

To handle this problem, recent method strides in Cross-Domain Recommendation (CDR) \cite{hu2018conet, yuan2019darec, fu2019deeply, yuan2020parameter, yang2021autoft, hande2021domain}. These algorithms commonly exploit the information from a source domain, characterized by abundant interactions relative to a target domain to extract domain-shareable information. Some approaches concentrate on duplicate users across both domains \cite{yuan2019darec, li2020ddtcdr, liu2021collaborative, guo2021gcn}. However, it is essential to note that these user-binding strategies may face constraints arising from the absence of overlapping (duplicate) users \cite{kang2019semi}. Alternatively, more flexible methods that operate independent of specific users or contexts have been introduced \cite{cai2019learning, zhao2020catn, krishnan2020transfer}. This trend has prompted the development of disentangled representation learning techniques \cite{gretton2005measuring, bousmalis2016domain, li2019learning, peng2019domain}, which can concurrently extract both domain-specific and domain-shareable knowledge. More recently, novel approaches contrived on review-based disentangled representation learning \cite{cao2022disencdr, choi2022based} free of duplicate users or contexts have demonstrated state-of-the-art performance.

Nonetheless, the above methods rely on Euclidean geometry for embedding \cite{globerson2004euclidean, khoshneshin2010collaborative}, which has been shown to be inadequate for modeling user-item bipartite graphs whose node degrees follow a power-law distribution \cite{clauset2009power}. This inadequacy leads to distortions in the embedding due to the exponential growth in the graph's volume with its radius. In addition, the application of hyperbolic geometry to cross-domain recommendation (CDR) remains relatively unexplored \cite{xu2023decoupled}, whereas recent studies on hyperbolic recommendation \cite{gulcehre2018hyperbolic, khrulkov2020hyperbolic,feng2020hme,vinh2020hyperml} focus on a single domain.

In this paper, we propose a novel strategy for hyperbolic CDR since previous methods rely on features extracted from two closely related domains \cite{cao2022disencdr, zhao2023crossa, zhao2023crossb}, potentially losing hierarchical information. For example, directly reducing the distance between two items, $i_1$ and $u_1$ (Figure \ref{euc_hyp}), purchased by a user without considering their positions can negatively impact the overall representation. As illustrated, distance-based minimization can lead to hierarchical collapse, as node positions cannot be preserved in this process. Specifically, if the degree of node $i_1$ is greater than others, placing a node with fewer interactions $u_1$ closer to the origin decreases the advantages of an exponentially increasing space. This context highlights the need for clever mechanisms that facilitate knowledge transfer while preserving a tree-like structure. To address this, we propose a solution with two strategies: degree-based normalization and structure alignment, substantiated through theoretical insights and empirical evidence from various experiments. In summary, our contributions are outlined as follows:
\begin{itemize}
\item We propose a novel CDR algorithm, called  \underline{H}yperbolic \underline{E}mbedding and Hierarchy-\underline{A}ware \underline{D}omain Disentanglement (HEAD), which enhances the previous review-based domain disentanglement by incorporating hyperbolic geometry. 
\item We propose degree-based hierarchy alignment and scale adjustment to enhance the knowledge transfer between two domains. To our knowledge, this is the first attempt to achieve domain disentanglement in a hyperbolic space.
\item We present theoretical understandings to prove the importance of hierarchy preservation in domain disentanglement.
\item We conducted various experiments to verify the effectiveness of our method and the accuracy of the theoretical analysis.
\end{itemize}

\section{Related Work}
In this section, we introduce the advent of recommender algorithms categorizing them as follows; (1) Review-based recommendations (using side information), (2) Cross-domain recommendations (knowledge transfer), and (3) Hyperbolic recommendations that can preserve hierarchical information.

\subsection{Review-Based Recommendation}
The explosive progress in text convolution techniques has ignited great attention in review-based recommender systems \cite{zheng2017joint, chen2019dynamic, chen2019co, dong2020asymmetrical}. For example, DeepCoNN \cite{zheng2017joint} employs two parallel convolutional neural networks (CNNs), and others further utilize an attention mechanism to exploit important words \cite{seo2017interpretable, chen2018neural, tay2018multi, dong2020asymmetrical}. Although these methods highlight the importance of review texts, their major limitation lies in the confined scale of target domains and the transfer of noisy data \cite{sachdeva2020useful, zeng2021zero}. To address these issues, cross-domain recommendation and disentanglement techniques have emerged to acquire useful information from richer domains.

\subsection{Cross-Domain Recommendation}
Cross-domain recommendation (CDR) utilizes additional information in extra (source) domain to address the sparseness in a target. Generally, these methods capture latent information from rating matrices or review texts \cite{fu2019deeply, krishnan2020transfer} to capture and transfer knowledge authored by overlapping users in both source and target domains \cite{elkahky2015multi, man2017cross, zhu2021cross}. Though certain methods \cite{wang2018cross, zhao2020catn} underscore the significance of employing non-overlapping users for generalization, they lack examination of what information would be efficiently transferred. Thus, recent studies have shifted focus towards identifying the most relevant aspects between two domains, commonly referred to as domain-shareable features. The foundational mechanism often commences with domain adaptation \cite{ramakrishnan2018overcoming, yuan2019darec, bonab2021cross,cao2022cross,chen2023knowledge}, which captures domain-shareable features through adversarial training. Advanced techniques that extract domain-specific and domain-shareable features simultaneously have been introduced more recently, collectively known as disentangled representation learning. These include MMT \cite{krishnan2020transfer}, DADA \cite{peng2019domain}, DisenCDR \cite{cao2022disencdr}, and SER \cite{choi2022based}. The pivotal aspect of this approach lies in domain disentanglement which strives to identify useful information for knowledge transfer.

\subsection{Hyperbolic Recommendation}
Most prior research on recommendation systems has primarily been conducted in an Euclidean space. These methods have demonstrated decent performance, but a challenge has been raised regarding their adequacy in modeling hierarchical structures such as user-item interactions or word vectors \cite{tifrea2018poincar}. Recent studies \cite{feng2020hme, wang2021hypersorec, su2023enhancing} utilize a hyperbolic space for representation learning, including informative collaborative filtering \cite{yang2022hicf, li2022hyperbolic, wang2023hdnr} with geometric regularization \cite{yang2022hrcf}. However, most of them utilize a single-domain dataset, which is susceptible to data sparsity. To address this problem, several researchers have integrated CDR with hyperbolic space embedding, but they disregard domain disentanglement \cite{xu2023decoupled, guo2023hyperbolic} and hierarchy structure preservation \cite{zhang2022geometric}, which are particularly crucial for accurate knowledge transfer in an exponentially expanded space. Thus, we aim to suggest a new hyperbolic CDR that preserves the structural property to better embed the user-item interactions.

\section{Preliminaries}
We start with the basic concepts of manifolds and hyperbolic geometry. Differential geometry defines three space types: hyperbolic, Euclidean, and spherical, based on curvatures. Especially, a hyperbolic space is one type of non-Euclidean space, which has a constant negative curvature at all points. In the literature, various mathematical formulations can be utilized to describe hyperbolic spaces, such as the Riemannian manifold \cite{lin2008riemannian}, Poincaré ball \cite{nickel2017poincare}, and Lorentz model \cite{nickel2018learning}. We employ the Poincaré ball for visualization \cite{gulcehre2018hyperbolic} and the Lorentz model for numerical operation \cite{law2019lorentzian}, respectively.

\textbf{Poincaré ball.} The representation of this model can be defined as $\mathcal{P}^d=(\mathcal{B}^d,g^{\mathcal{B}}_x)$, which stands for the open $n$-dimensional unit ball $\mathcal{B}^d=\{x \in \mathcal{R}^d: k||x|| < 1\}$ and hyperbolic feature $g^{\mathcal{B}}_x$:
\begin{equation}
\label{poin_ball}
    g^{\mathcal{B}}_x = \mathrm{({2 \over 1-k||x||^2})}^2g^E_x,
\end{equation}
where $k$ is the radius of the ball. The above equation converts an Euclidean metric tensor $g^E$ to a hyperbolic one. If $k=0$, we can easily infer that the ball is identical to the Euclidean space. Also, a distance function on $\mathcal{P}$ is defined as below:
\begin{equation}
\label{poincare_dist}
    \mathrm{d_{\mathcal{P}}(x,y)=\sqrt{k} \ arcosh\left(1+2k{||x-y||^2 \over (k-||x||^2)(k-||y||^2)}\right)}
\end{equation}

\textbf{Lorentz model.} Similarly, the Lorentz model is defined as $\mathcal{L}^d=(\mathcal{H}^d,g^{\mathcal{H}}_x)$, where $\mathcal{H}^d=\{x \in \mathcal{R}^{d+1}:<x,x>_\mathcal{L}=-k,x_0>0\}$. Here, $<,>_{\mathcal{L}}$ is the Lorentizan inner product:
\begin{equation}
    \mathrm{<x,y>_{\mathcal{L}}=-x_0y_0+\sum^n_{i=1}x_iy_i},
\end{equation}
and $g^{\mathcal{H}}_x=diag(-1,1,...,1)$ is a positive-definite metric tensor to calculate a distance of two points $x,y \in \mathcal{H}^d$ as follows:
\begin{equation}
\label{lorentz_dist}
    \mathrm{d_{\mathcal{L}}(x,y)=\sqrt{k} \ arcosh(-{<x,y>_\mathcal{L} \over k})}
\end{equation}
For a certain point $x \in \mathcal{H}^d$ in hyperbolic space, we can define the tangent space centered at $x$ as below: 
\begin{equation}
    \mathrm{\mathcal{T}_x\mathcal{H}^d=\{v \in \mathcal{R}^{d+1}: <v,x>_{\mathcal{L}}=0\}},
\end{equation}
where the orthogonality holds for all $v$ concerning the Lorentz scalar product. Using these characteristics, conversions between the tangent and hyperbolic space can be achieved through exponential and logarithmic maps as follows.
\begin{itemize}
    \item (Exponential map) $\mathcal{T}_x\mathcal{H}^d \rightarrow \mathcal{H}^d$ projects $v$ onto hyperbolic space as,
    \begin{equation}
    \label{exp_map}
        \mathrm{exp_x(v)=cosh({||v||_{\mathcal{L}} \over \sqrt{k}})x + \sqrt{k}sinh({||v||_{\mathcal{L}} \over \sqrt{k}}){v \over ||v||_{\mathcal{L}}}}
    \end{equation}
    \item (Logarithmic map) $\mathcal{H}^d \rightarrow \mathcal{T}_x\mathcal{H}^d$ projects $v$ back to Euclidean space as,
    \begin{equation}
    \label{log_map}
        \mathrm{log_x(v)=d_{\mathcal{L}}(x,v){v+{1 \over k}<x,v>_{\mathcal{L}}x \over ||v+{1 \over k}<x,v>_{\mathcal{L}}x||}_{\mathcal{L}}}
    \end{equation}
\end{itemize}


\section{Methodology}
\begin{figure}[t]
  \includegraphics[width=.49\textwidth]{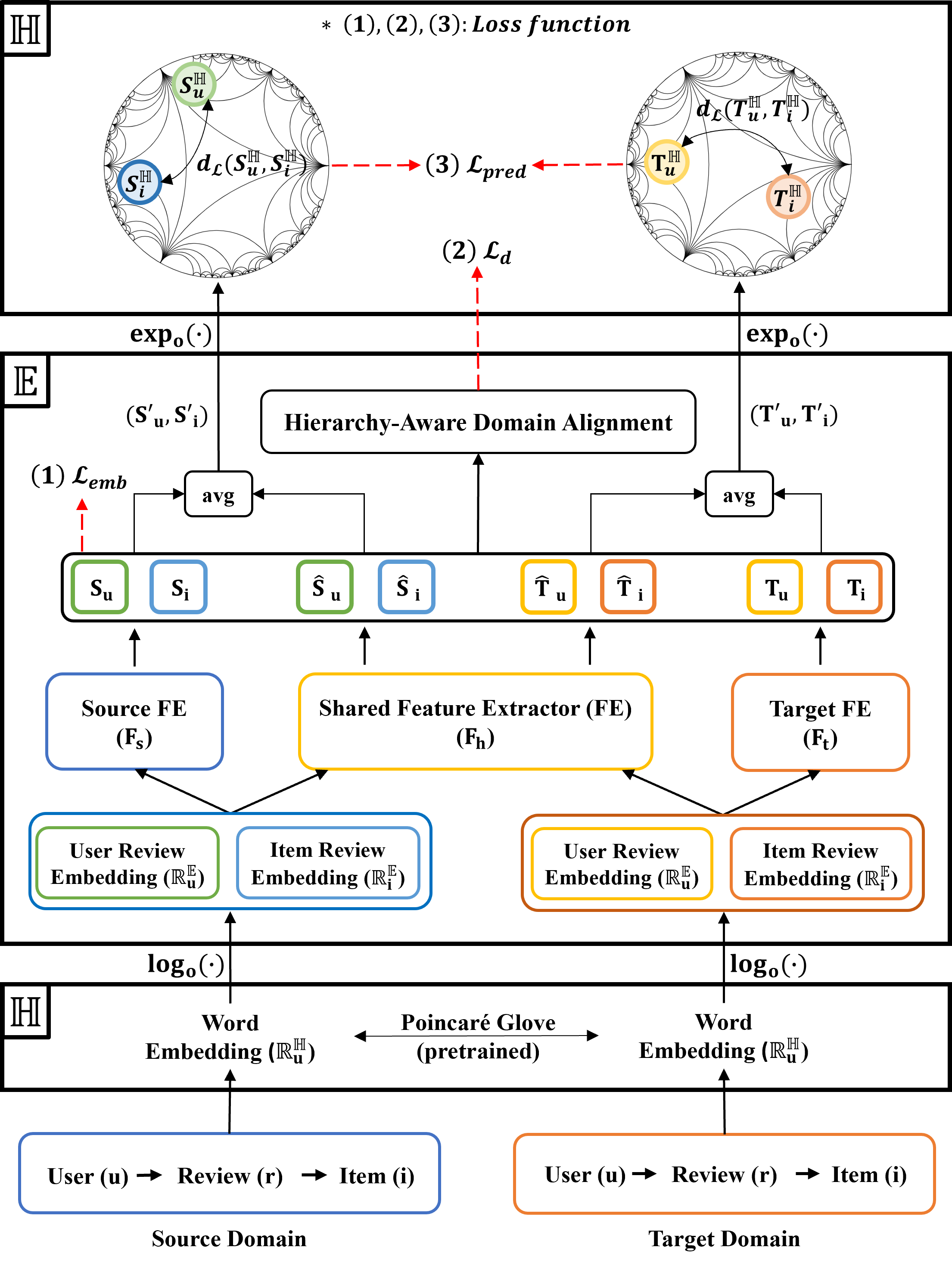}
  \caption{The overall framework of the Hierarchy-Aware Hyperbolic Embedding and Domain Disentanglement (HEAD) scheme. The (1)-(3) represents three types of loss functions} 
  \label{model}
\end{figure}

In Figure \ref{model}, we illustrate the overall architecture of our model, called HEAD (\underline{H}yperbolic \underline{E}mbedding and Hierarchy-\underline{A}ware \underline{D}omain Disentanglement, which is comprised of the following key components:

\begin{itemize}
    \item \textbf{Word embedding.} This part vectorizes reviews using pre-trained word embedding. In contrast to the previous methods that adopt Euclidean embedding such as word2vec\footnote{https://code.google.com/archive/p/word2vec} \cite{mikolov2013distributed} or Euclidean GloVe\footnote{https://nlp.stanford.edu/projects/glove} \cite{pennington2014glove}, we employ the Poincaré Glove\footnote{\url{https://github.com/alex-tifrea/poincare_glove}} \cite{tifrea2018poincar} which better preserves the hierarchical property of words.
    \item \textbf{Feature extraction.} This module elicits pertinent information from embedded documents using three types of feature extractors (FEs) \cite{choi2022based}; the shared FE focuses on domain-shareable knowledge for transfer while the source and target FEs capture the domain-specific features.
    \item \textbf{Hierarchy-aware embedding and domain disentanglement.} Extracted features are aligned hierarchically and then knowledge is transferred while retaining this structure. This also reinforces the separability of domain discriminator.
    \item \textbf{Prediction and optimization.} Outputs are integrated and projected back to the hyperbolic space for prediction. A marginal ranking loss is computed for optimization.
\end{itemize}

\subsection{\textbf{Word Embedding}}
Each data record follows a format (\textit{u, i, $y_{u,i}$, $r_{u,i}$}), which means that a user $u$ purchased an item $i$ and left a rating $y_{u,i}$ and a review $r_{u,i}$. Assume that we are concerned with a possibly unseen rating $y_{u,i}$. We first aggregate all reviews of $u$ and $i$. Specifically, for a user $u$, we gather all reviews written by her except for the specific pair $r_{u,i}$ (not available during inference) and consider them as a single document $R_{u}$. Likewise, one can construct the collection of item reviews, $R_{i}$. Here, we ignore the temporal sequence of ratings or reviews. Finally, we apply the word embedding function to $R_{u}$ and $R_{i}$. Although many strategies are applicable (e.g, word2vec\footnote{https://code.google.com/archive/p/word2vec} \cite{mikolov2013distributed}), we focus on the Euclidean- \cite{pennington2014glove} and Poincaré Glove \cite{tifrea2018poincar} here. 
\begin{itemize}
    \item \textbf{(Euclidean Glove)} increases the co-occurrence probability ($X_{ij}$) of the central word ($w_i$) and its neighbor words ($\tilde{w}_j$) as,
    \begin{equation}
        \mathrm{\min_w\sum^V_{i,j}f(X_{ij})(w^T_i\tilde{w}_j+b_i+\tilde{b}_j-logX_{ij})^2},
    \end{equation}
    where $b$ stands for the bias term.
    \item \textbf{(Poincaré Glove)} replaces $w^T_i\tilde{w}_j$ with $-h(d_{\mathcal{P}}(w^T_i,\tilde{w}_j))$ as,
    \begin{equation}
        \mathrm{\min_w\sum^V_{i,j}f(X_{ij})(-h(d_{\mathcal{P}}(w^T_i,\tilde{w}_j))+b_i+\tilde{b}_j-logX_{ij})^2},
    \end{equation}
where $d_{\mathcal{P}}$ is a distance function in Eq. \ref{poincare_dist} and $h(x)=cosh^2(x)$. 
\end{itemize}
The Euclidean and Poincaré renditions \footnote{\url{https://polybox.ethz.ch/index.php/s/TzX6cXGqCX5KvAn}} have been trained using 1.4 billion tokens from English Wikipedia, and we compare them as pre-trained word embedding $f(\cdot)$ in Table \ref{performance_ndcg}. Using this, the textual documents $R_{u}$ and $R_{i}$ are mapped into the matrix $\mathrm{R^{\mathbb{E}}_u=f(R_{u}), R^{\mathbb{E}}_i=f(R_{i})}$ of $\mathcal{R}^{n \times d}$, where $n$ is the vocabulary size and $d$ is the embedding dimension. We project these matrices onto the hyperbolic space using the exponential map in Eq. \ref{exp_map} as below:
\begin{equation}
\begin{gathered}
    \mathrm{R^{\mathbb{H}}_{u}=exp_o(R^{\mathbb{E}}_{u})=[cosh(||R^{\mathbb{E}}_{u}||), sinh(||R^{\mathbb{E}}_{u}||){R^{\mathbb{E}}_{u} \over ||R^{\mathbb{E}}_{u}||}]} \\
    \mathrm{R^{\mathbb{H}}_{i}=exp_o(R^{\mathbb{E}}_{i})=[cosh(||R^{\mathbb{E}}_{i}||), sinh(||R^{\mathbb{E}}_{i}||){R^{\mathbb{E}}_{i} \over ||R^{\mathbb{E}}_{i}||}]}
\end{gathered}
\end{equation}
We set the curvature $k=1$ in Eq. \ref{exp_map} for simplicity.

\subsection{\textbf{Feature Extraction}}
The above word embedding procedure is applied to the source and target domains, respectively.
Given word embedding $R^{\mathbb{H}}_{u}$ and $R^{\mathbb{H}}_{i}$ from each domain, we aim to extract useful information. Several feature extraction strategies have been proposed, including domain adaptation \cite{yuan2019darec}, variational reconstruction \cite{liu2022exploiting}, personalized transfer \cite{zhu2022personalized}, contrastive learning \cite{xie2022contrastive}, and domain disentanglement \cite{peng2019domain}. Among them, we adopt the domain disentanglement algorithm, which does not obligate user overlapping in both domains \cite{cao2022disencdr}. For this, as illustrated in Figure \ref{model}, we employ three types of feature extractors (FEs), all of which consist of simple multi-channel Convolutional Neural Networks (CNNs). Specifically, the shared FE processes datasets from both domains, while the source and target FEs deal with the documents from their domains only (please refer to \cite{choi2022based} for more details). For the sake of simplicity, we focus on the mechanisms in the source domain since the target domain procedure is the same. The feature extraction process is given by:
\begin{equation}
\begin{gathered}
\label{convolution}    \mathrm{S_u=F_s(\log_o(R^{\mathbb{H}}_u)), \,\, S_i= F_s(\log_o(R^{\mathbb{H}}_i))} \\
    \mathrm{\widehat{S}_u=F_h(\log_o(R^{\mathbb{H}}_u)), \,\, \widehat{S}_i= F_h(\log_o(R^{\mathbb{H}}_i))} \\
\end{gathered}
\end{equation}
As illustrated, the hyperbolic embedding of the user and item is projected back to Euclidean space using a logarithmic map in Equation \ref{log_map}, followed by the application of CNNs ($F_s$ and $F_h$). Consequently, the source domain yields four outputs from two feature extractors, denoted as ${S_u, S_i, \widehat{S}_u, \widehat{S}_i}$ (depicted in the middle of Figure \ref{model}). Additional insights into CNNs can be found in \cite{zheng2017joint}.

\textbf{Remark.} The large language models (e.g., LLaMA \cite{touvron2023llama}, ChatGPT-4 \cite{achiam2023gpt}) may replace CNNs if their parameters can be fine-tuned.

\subsection{Hierarchy-Aware Hyperbolic Embedding and Domain Disentanglement}
We propose two constraints to achieve domain disentanglement between the extracted features while preserving the hierarchy.

\subsubsection{\textbf{Hierarchy-aware hyperbolic embedding.}}
Recent work \cite{khrulkov2020hyperbolic} reveals that the uncertainty decreases as the embedding gets closer to the boundary of the Poincaré ball. HRCF \cite{yang2022hrcf} suggests a hyperbolic regularization optimized for the characteristics of a power-law distribution. Specifically, users or items with many interactions (dense) are pulled to the center, while sparse ones are placed near the boundary. For this, HRCF identifies the root as the average of the entire embedding to make it as an origin below:
\begin{equation}
\label{s_root}
    \mathrm{S_u^{root}={1 \over N_u}\sum^{N_u}_{u'=1}{1 \over 2}(S^{u'}_u+\widehat{S}^{u'}_u}), \,\,\, \mathrm{S_i^{root}={1 \over N_i}\sum^{N_i}_{i'=1}{1 \over 2}(S^{i'}_i+\widehat{S}^{i'}_i})
\end{equation}
The $N_u$ and $N_i$ are the total number of users and items, respectively. Then, they apply so-called root alignment that locates the nodes to be separated from the origin as below:
\begin{equation}
\label{hyp_align}
    \mathrm{S_u^{norm}={1 \over N_u}\sum^{N_u}_{u'=1}||S_u^{u'}-S_u^{root}||^2_2, \,\,\, S_i^{norm}={1 \over N_i}\sum^{N_i}_{i'=1}||S_i^{i'}-S_i^{root}||^2_2}
\end{equation}
The loss function is the inverse of the above equation, which aims to decrease the central density. However, this strategy might not sufficiently reflect the popularity of nodes since it simply pushes all nodes away from the center. To alleviate this problem, we suggest to modify Eq. \ref{hyp_align} as follows.
\begin{manualprop}{4.1}[Hierarchy-aware hyperbolic embedding]\label{prop_hierarchy} The loss function is normalized based on the maximum node degree, $\max(d)$, as follows: 
\begin{flalign}
    \label{hie_emb}
    \mathrm{S_{u,deg}^{norm}}&=\mathrm{{1 \over N_u}\sum^{N_u}_{u'=1}{\max(d_u)-d_{u'} \over \max(d_u)}||S_u^{u'}-S_u^{root}||^2_2} \\
    \mathrm{S_{i,deg}^{norm}}&=\mathrm{{1 \over N_i}\sum^{N_i}_{i'=1}{\max(d_i)-d_{i'} \over \max(d_i)}||S_i^{i'}-S_i^{root}||^2_2}
\end{flalign}
Notation $d_{u'}$ and $d_{i'}$ stands for the degrees of user $u'$ and item $i'$, respectively. Through this, we can place popular users and items near the origin, while pushing low-degree nodes towards the boundary.
\end{manualprop}
\textup{\textit{\textbf{Proof}}. see proof of proposition \ref{thm_deg} in Section \ref{theo_anal}.}

The loss in the target domain can be retrieved similarly. Then, we can define the hierarchical embedding loss as below:
\begin{equation}
\label{emb_loss}
\mathrm{\mathcal{L}_{emb}=1/\sqrt{S_{u,deg}^{norm}+S_{i,deg}^{norm}+T_{u,deg}^{norm}+T_{i,deg}^{norm}}}
\end{equation}

\subsubsection{\textbf{Hierarchy-aware domain disentanglement.}}
Knowledge transfer has gained substantial attention in the field of cross-domain recommendation. In this regard, we align with the recently proposed domain disentanglement algorithm \cite{cao2022disencdr, choi2022based}, which operates without the requirement of overlapping users. The fundamental mechanism of domain disentanglement can be delineated as follows: (1) domain-specific features $S,T$ should be readily inferable to their originating domains to mitigate domain discrepancies, and (2) domain-shareable features $\widehat{S},\widehat{T}$ ought to encapsulate domain-indiscriminative information, ensuring pairwise independence between domains \cite{gretton2005measuring, li2019learning, nema2021disentangling}. For this, we first concatenate the user and item vectors from each feature extractor as below:
\begin{equation}
\begin{gathered}
    \mathrm{S=[S_u \oplus S_i], \,\,\, T=[T_u \oplus T_i]} \\
    \mathrm{\tilde{S}=[\widehat{S}_u \oplus \widehat{S}_i], \,\,\, \tilde{T}=[\widehat{T}_u \oplus \widehat{T}_i]}
\end{gathered}
\end{equation}
Before delving into the scale alignment module, we introduce the following discussion on previous disentanglement algorithms. Prior methods of domain disentanglement, such as those proposed by \cite{peng2019domain,krishnan2020transfer,cao2022disencdr,choi2022based,zhang2022geometric} merely focus on preserving the scale of the extracted features. In detail, they simply forward these features to the domain discriminator ($F_d$) in the following manner:
\begin{equation}
\begin{gathered}
    d_S=F_d(S), \,\,\, d_T=F_d(T) \\
    \tilde{d}_S=F_d(g(\tilde{S})), \,\,\, \tilde{d}_T=F_d(g(\tilde{T}))
\end{gathered}
\label{dom_shr}
\end{equation}

The $F_d$ consists of the two layers of a fully connected neural network and the notation $d_S, d_T \in \{0,1\}$ denotes the predicted domain (0/1 are the source/target). Additionally, $g(\cdot)$ signifies the Gradient Reversal Layer (GRL), which remains inactive during the forward propagation but reverses the sign of the gradient during the back-propagation (for detailed information, refer to \cite{mansour2009domain, peng2019domain}). A major limitation of these methods lies in the disruptive changes in positional information. As elucidated earlier, domain-shareable features tend to converge, while domain-specific features separate apart. Consequently, minor positional changes can cause exponentially magnified effects. A similar issue is observed in other domain alignment methods. To solve the problem, some directly reduce the distance of the same set of users between domains \cite{zhao2023crossb}, and others leverage variational inference to align the mean and variance of feature distributions \cite{liu2022exploiting}. In this paper, we propose a novel disentanglement strategy that conserves the scale of the extracted information by revising Eq. \ref{dom_shr} as follows.

\begin{manualprop}{4.2}[Scale Alignment]\label{}
    We adjust the scale of inputs before applying the domain discriminator as follows:
    \begin{itemize}
        \item (Scale alignment between domain-specific knowledge)
        \begin{equation} 
        \begin{gathered}
        d_S=F_d({S \over |S|}), \, \,d_T=F_d({T \over |T|})
        \end{gathered}
        \label{hie_dom_spe}
        \end{equation}
        \item (Scale alignment between domain-shareable knowledge)
        \begin{equation} 
        \begin{gathered}
        \tilde{d}_S=F_d(g({\tilde{S} \over |\tilde{S}|})),\, \,\tilde{d}_T=F_d(g({\tilde{T} \over |\tilde{T}|}))
        \end{gathered}
        \label{hie_dom_shr}
        \end{equation}
    \end{itemize}
Based on this, we can define the domain loss $\mathcal{L}_d$ as,
\begin{flalign}
\label{da_loss}
    \mathcal{L}_d= &-{1 \over N_s}\sum_{n=1}^{N_s} log(1-d_S) -{1 \over N_s}\sum_{n=1}^{N_s} log(1-\tilde{d}_S) \\
    &-{1 \over N_t}\sum_{n=1}^{N_t} log(d_T)-{1 \over N_t}\sum_{n=1}^{N_t} log(\tilde{d}_T) \nonumber
\end{flalign}
\end{manualprop}
\textup{\textit{\textbf{Proof}}. see proof of proposition \ref{thm_str} in Section \ref{theo_anal}}. \\

Finally, we claim that scale alignment can also enhance the separability of a discriminator in the proposition below. 
\begin{manualprop}{4.3}[Advantages of Scale adjustment] 
    Removing the scale of input features enhances the domain discriminator's separability and guarantees stable convergence.
    
    \textup{\textit{\textbf{Proof}}. see proof of proposition \ref{thm_dis} in Section \ref{theo_anal}}.
\end{manualprop}

\subsection{Inference and Optimization}
\label{sec_opt}
\textbf{Inference.} We aim to measure the relativity between a user and an item using their aggregated features. To elaborate, in the middle of Figure \ref{model} (source domain), we compute the average (avg) of the outputs from the source ($S$) and shared FEs ($\widehat{S}$), adding the latent of user-item interaction vectors ($p_u$ and $p_i$) as follows:
\begin{equation}
    S'_u = {1 \over 2}(S_u + \widehat{S}_u)+p_u, \,\,\, S'_i = {1 \over 2}(S_i + \widehat{S}_i)+p_i
\end{equation}
Then, we project the aggregated representation onto the hyperbolic space using the exponential map in Eq. \ref{exp_map} as follows:
\begin{flalign}
    \mathrm{S^{\mathbb{H}}_u}&\mathrm{=exp_o(S'_u)=(cosh(||S'_u||), sinh(||S'_u||){S'_u \over ||S'_u||})} \\
    \mathrm{S^{\mathbb{H}}_i}&=\mathrm{exp_o(S'_i)=(cosh(||S'_i||), sinh(||S'_i||){S'_i \over ||S'_i||})}
\end{flalign}
Finally, we can measure their distance as below:
\begin{equation}
    p(S^{\mathbb{H}}_u,S^{\mathbb{H}}_i)=\mathcal{M}(S^{\mathbb{H}}_u \oplus S^{\mathbb{H}}_i) \, d_{\mathcal{L}}(S^{\mathbb{H}}_u, S^{\mathbb{H}}_i),
\end{equation}
where $\mathcal{M}(\cdot) \in [0,1]$ is a MLP with Sigmoid activation function. This adjusts the distance between users and items, $d_{\mathcal{L}}(\cdot)$ (Eq. \ref{lorentz_dist}) to reflect the user's preference for popular items. For optimization, we adopt the hyperbolic margin ($\epsilon=0.1$) ranking loss \cite{sun2021hgcf} given the positive ($i$) and negative sample ($j$) as,
\begin{equation}
\label{pred_loss}
\mathcal{L}_{pred}=\max(p(S^{\mathbb{H}}_u,S^{\mathbb{H}}_i)^2-p(S^{\mathbb{H}}_u,S^{\mathbb{H}}_j)^2+\epsilon, 0)
\end{equation}

\textbf{Optimization.} We define the overall objective function as to minimize the weighted sum of Eq. \ref{emb_loss}, \ref{da_loss}, \ref{pred_loss} as below:
\begin{equation}
\label{total_loss}
\min_{\theta}\mathcal{L}_{total} = \lambda_1 \mathcal{L}_{emb} + \lambda_2 \mathcal{L}_{d} + \mathcal{L}_{pred} + \delta ||\theta||
\end{equation}
The hyperparameters $\lambda_1$ and $\lambda_2$ balance the losses. For each dataset, we find $\lambda_1$ and $\lambda_2$ through grid search that yielded the best validation score (Fig. \ref{param_sense}). The parameter $\theta$ is optimized using the Adam optimizer, with $\delta$ representing the regularization term. Additionally, we practiced early stopping within 300 iterations and applied negative sampling for ratings that meet the condition of $y_{u,j} \leq 3$.

\subsection{Time Complexity}
In addition to the plain text convolution module ($A$), we employ a hyperbolic Glove that requires a mapping from the hyperbolic space to the Euclidean ones ($B$). Secondly, the discriminator is a simple two-layer neural network and the degree normalization only averages the outputs ($C$). Lastly, the scale alignment has linear complexity as it only matches the magnitudes of the two vectors ($D$). Thus, the complexity is $\mathcal{O}((A+B+C+D) \cdot N_t) \approx \mathcal{O}(N_t)$, which is a linear model proportional to the size of a target domain.

\subsection{Theoretical Analysis} \label{theo_anal}
\begin{manualtheorem}{4.1}[Nodes with smaller degrees are likely to be pushed away from the origin]\label{thm_deg}
Let us take $S^{norm}_{u,deg}$ (Eq. \ref{hie_emb}) as an example. Since we minimize the $\mathcal{L}_{emb}$ in Eq. \ref{emb_loss}, the parameter $F_s$ in Eq. \ref{convolution} is trained to maximize $S^{norm}_u$ as follows:
\begin{equation}
\label{eq_prop4.1}
    \underset{F_s}{\arg\max}\,S^{norm}_{u,deg}={\max(d)-d_u \over \max(d)}{\partial \mathcal{L}_{emb} \over \partial F_s}
\end{equation}
Both our proposed method (Eq. \ref{hie_emb}) and the plain method (Eq. \ref{hyp_align}) share the second term in Eq. \ref{eq_prop4.1}. Thus, we focus on the first term (degree normalization) that determines the scale of gradient as below:
\begin{equation}
    ||\bigtriangledown_{F_s} S^{norm}_{u,deg} \,|| \,\, / \,\, ||\bigtriangledown_{F_s} S^{norm}_u \,|| \approx {\max(d)-d_u \over \max(d)},
\end{equation}
where $\bigtriangledown$ denotes the partial derivative. Since $(\max(d)-d_u ) / \max(d) \in [0, 1]$, the scale of gradient increases as the degree of nodes ($d_u$) decreases, pushing it away from the origin and vice versa.
\end{manualtheorem}

\begin{manualtheorem}{4.2}[Scale Preservation]\label{thm_str}
Let us take two domain-specific features $d_S$ and $d_T$. According to the law of cosines, the following equality holds:
\begin{equation}
\label{plane_dist}
||d_S-d_T||^2=||d_S||^2+||d_T||^2-2||d_S||\cdot||d_T||\cos C,
\end{equation}
where $C$ is the angle between the vectors. Since they are from the domain-specific FEs, the updated features ($d'_S,d'_T$) satisfy $||d'_S-d'_T||^2 > ||d_S-d_T||^2$. Thus, we can redefine the Eq. \ref{plane_dist} as below:
\begin{equation}
||d'_S||^2+||d'_T||^2-2||d'_S||\cdot||d'_T||\cos C' > ||d_S||^2+||d_T||^2-2||d_S||\cdot||d_T||\cos C
\end{equation}
Since $\cos C' < \cos C$, assuming the update function as $d'_S=d_S-\bigtriangledown_{d_S}\mathcal{L}_d$, we can infer that the scale increases in proportion to $||d_S||$. However, our method in Eq. \ref{hie_dom_shr} can preserve the scale because $d'_S=d_S / ||d_S|| - \bigtriangledown_{d_S / ||d_S||} \mathcal{L}_d$, which is proportional to $d_S / ||d_S|| \approx 1$.
\end{manualtheorem}

\begin{manualtheorem}{4.3}[Scale adjustment enhances stability and domain separability]\label{thm_dis}
The classification error is associated with the distance from the decision boundary \cite{yan2022two} or the distance between two feature vectors \cite{furusho2019resnet}. Let the weight matrix of the domain discriminator be $W$, the activation function be $\phi$. Given two inputs $x$ and $y$, the separability $\mathcal{S}$ can be defined by an inner product, reflecting both the scale and the angle, as follows:
\begin{equation}
    \mathcal{S}=\phi(Wx)^T\phi(Wy)=||x|| \cdot ||y||\cdot f(x,y)
\end{equation}
The $f(x,y)$ is the angle (e.g., cosine similarity) between two vectors. Thus, the partial derivative of the separability is given by:
\begin{equation}
    \bigtriangledown_W\mathcal{S}=||x|| \cdot ||y||\cdot {\partial \mathcal{S} \over \partial W}f(x,y)
\end{equation}
Here, we focus on the scale of the gradient. Since the angle lies in $-1 \leq f(x,y) \leq 1$, the gradient $\bigtriangledown_W\mathcal{S}$ depends on the scale of two inputs, and removing this information is considered as one type of feature scaling method \cite{wan2019influence,chen2022feature}. Thus, we can guarantee stable convergence without being affected by input's covariations, $cov(x,y) \approx ||x|| \cdot ||y||$ as the following inequality holds, $0 \leq \bigtriangledown_W\mathcal{S} \, / \, (||x|| \cdot ||y||) \leq 1$.
\end{manualtheorem}

\section{Experiments}
We set fundamental questions to provide a comprehensive analysis of the proposed method. The details of the following research questions (RQs) are explained from Section \ref{rq1} to Section \ref{rq4}:
\begin{itemize}
\item \textbf{RQ1:} Does our model achieve a significant performance improvement compared to state-of-the-art baselines?
\item \textbf{RQ2:} In addition to the experimental results, does scale alignment enhance domain disentanglement?
\item \textbf{RQ3:} Does HEAD preserve the hierarchical structure better than previous methods?
\item \textbf{RQ4:} How sensitive is the performance of proposed method on hyperparameters $\lambda_1$ and $\lambda_2$ in Eq. \ref{total_loss}?
\end{itemize}

\begin{table}
\caption{Details of the benchmark datasets}
\label{dataset}
\centering
\begin{adjustbox}{width=0.48\textwidth}
\begin{tabular}{c|llll}
\multicolumn{1}{l}{}    & \multicolumn{1}{l}{}       &         &         &                \\ 
\Xhline{2\arrayrulewidth}
\multicolumn{1}{c|}{Domain}        & Dataset                       & \# users & \# items & \# reviews  \\ 
\Xhline{2\arrayrulewidth}
\multirow{3}{*}{Source} 
                        & Clothing (Cloth)                     & 1,219,520 & 376,858  & 11,285,464        \\
                        & CDs and Vinyl (CDs)               & 112,391 & 73,713 & 1,443,755       \\
                        & Toys and Games (Toys)             & 208,143   & 78,772   & 1,828,971        \\
\hline
\multirow{4}{*}{Target} & Luxury Beauty            & 3,818   & 1,581   & 34,278         \\
                        & All Beauty             & 990 & 85 & 5,269         \\
                        & Digital Music                 & 16,561   & 11,797    & 169,781         \\
                        & Video Games      & 55,217   & 17,408     & 497,577         \\
\Xhline{2\arrayrulewidth}
\end{tabular}
\end{adjustbox}
\end{table}

\begin{table*}[!htbp]
\centering
\caption{ (RQ1) The performance on four target domain datasets with a significance level \textbf{*} ($\rho$-value$<0.05$). The \color{blue}blue\color{black} \, and \color{red}red\color{black} \, indicate best NDCG@10 (ND) and HR@10 (HR) scores. A symbol ($\mathbb{H}$) indicates that the method uses the hyperbolic space. The HEAD$\mathrm{^*_E}$ and HEAD$\mathrm{^*_P}$ employ the Euclidean- \cite{pennington2014glove} and Poincaré- \cite{tifrea2018poincar} Glove, respectively}
\label{performance_ndcg}
\begin{tabularx}{\textwidth}{@{}c|c|c|ccc|ccc|ccc|ccc@{}}
\cmidrule[2pt]{1-15}
\multirow{2}{*}{} & \multirow{2}{*}{Method} & \multirow{2}{*}{@10} &  \multicolumn{3}{c|}{\underline{Luxury Beauty}} & \multicolumn{3}{c|}{\underline{All Beauty}} & \multicolumn{3}{c|}{\underline{Digital Music}} & \multicolumn{3}{c}{\underline{Video Games}} \\
& &  & Cloth & CDs & Toys & Cloth & CDs & Toys & Cloth & CDs & Toys & Cloth & CDs & Toys \\
\cmidrule[1.5pt]{1-15}
\multirow{11}{*}{\rotatebox{90}{Single-Domain}} 
& \multirow{2}{*}{DeepCoNN} & ND & \multicolumn{3}{c|}{0.093} & \multicolumn{3}{c|}{0.089} & \multicolumn{3}{c|}{0.101} & \multicolumn{3}{c}{0.124} \\
& & HR &  \multicolumn{3}{c|}{0.177} & \multicolumn{3}{c|}{0.165} & \multicolumn{3}{c|}{0.184} & \multicolumn{3}{c}{0.220} \\ \cmidrule[.1pt]{2-15}
& \multirow{2}{*}{AHN} & ND &  \multicolumn{3}{c|}{0.129} & \multicolumn{3}{c|}{0.142} & \multicolumn{3}{c|}{0.106} & \multicolumn{3}{c}{0.171} \\ 
& & HR &  \multicolumn{3}{c|}{0.231} & \multicolumn{3}{c|}{0.254} & \multicolumn{3}{c|}{0.199} & \multicolumn{3}{c}{0.306} \\  \cmidrule[.1pt]{2-15}
 & \multirow{2}{*}{HGCF$^{\mathbb{H}}$} & ND &  \multicolumn{3}{c|}{0.123} & \multicolumn{3}{c|}{0.135} & \multicolumn{3}{c|}{0.121} & \multicolumn{3}{c}{0.168} \\
 &  & HR &  \multicolumn{3}{c|}{0.242} & \multicolumn{3}{c|}{0.250} & \multicolumn{3}{c|}{0.217} & \multicolumn{3}{c}{0.289} \\ \cmidrule[.1pt]{2-15}
 & \multirow{2}{*}{GDCF$^{\mathbb{H}}$} & ND &  \multicolumn{3}{c|}{0.121} & \multicolumn{3}{c|}{0.140} & \multicolumn{3}{c|}{0.118} & \multicolumn{3}{c}{0.153} \\
 &  & HR &  \multicolumn{3}{c|}{0.229} & \multicolumn{3}{c|}{0.251} & \multicolumn{3}{c|}{0.220} & \multicolumn{3}{c}{0.277} \\
 \cmidrule[.1pt]{2-15}
 & \multirow{2}{*}{HDNR$^{\mathbb{H}}$} & ND &  \multicolumn{3}{c|}{0.144} & \multicolumn{3}{c|}{0.148} & \multicolumn{3}{c|}{0.133} & \multicolumn{3}{c}{0.189} \\
 &  & HR &  \multicolumn{3}{c|}{0.262} & \multicolumn{3}{c|}{0.270} & \multicolumn{3}{c|}{0.251} & \multicolumn{3}{c}{0.344} \\
\cmidrule[1.5pt]{1-15}

\multirow{15}{*}{\rotatebox[origin=c]{90}{Cross-Domain}} 
& \multirow{2}{*}{DDTCDR} & ND & 0.072 & 0.054 & 0.059 & 0.054 & 0.045 & 0.041 & 0.065 & 0.079 & 0.069 & 0.062 & 0.071 & 0.083 \\
& & HR & 0.138 & 0.100 & 0.112 & 0.103 & 0.086 & 0.075 & 0.121 & 0.141 & 0.130 & 0.118 & 0.134 & 0.151 \\ \cmidrule[.1pt]{2-15}
& \multirow{2}{*}{RC-DFM} & ND & 0.137 & 0.114 & 0.122 & 0.135 & 0.132 & 0.128 & 0.103 & 0.118 & 0.115 & 0.131 & 0.135 & 0.146 \\
& & HR& 0.256 & 0.211 & 0.233 & 0.260 & 0.254 & 0.249 & 0.201 & 0.231 & 0.222 & 0.248 & 0.261 & 0.266 \\ \cmidrule[.1pt]{2-15}
& \multirow{2}{*}{CATN} & ND & 0.141 & 0.117 & 0.125 & 0.140 & 0.133 & 0.131 & 0.102 & 0.118 & 0.123 & 0.144 & 0.137 & 0.172 \\
& & HR& 0.271 & 0.218 & 0.237 & 0.258 & 0.259 & 0.251 & 0.198 & 0.224 & 0.221 & 0.240 & 0.263 & 0.302 \\ \cmidrule[.1pt]{2-15}
& \multirow{2}{*}{MMT} & ND & 0.146 & 0.125 & 0.139 & 0.142 & 0.136 & 0.136 & 0.117 & 0.130 & 0.122 & 0.161 & 0.156 & 0.188 \\
& & HR& 0.270 & 0.241 & 0.264 & 0.268 & 0.255 & 0.253 & 0.216 & 0.244 & 0.229 & 0.298 & 0.300 & 0.351 \\ \cmidrule[.1pt]{2-15}
& \multirow{2}{*}{SER} & ND & 0.149 & 0.136 & 0.147 & 0.150 & 0.143 & 0.146 & 0.149 & 0.152 & 0.148 & 0.199 & 0.205 & 0.221 \\
& & HR& 0.283 & 0.270 & 0.286 & 0.288 & 0.272 & 0.279 & 0.261 & 0.300 & 0.297 & 0.353 & 0.352 & 0.394 \\ \cmidrule[.1pt]{2-15}
& \multirow{2}{*}{DH-GAT$^{\mathbb{H}}$} & ND & 0.152 & 0.142 & 0.145 & 0.153 & 0.150 & 0.152 & 0.146 & 0.144 & 0.127 & 0.200 & 0.202 & 0.214 \\
 &  & HR& 0.285 & 0.266 & 0.271 & 0.290 & 0.289 & 0.294 & 0.278 & 0.278 & 0.246 & 0.366 & 0.371 & 0.389 \\
\cmidrule[1.5pt]{1-15}
\multirow{4.5}{*}{\rotatebox[origin=c]{90}{Ours}} & \multirow{2}{*}{HEAD$\mathrm{^{\mathbb{H}}_E}$} & ND & 0.162$^{*}$ & 0.160$^{*}$ & 0.163$^{*}$ & 0.154$^{*}$ & 0.150 & 0.153$^{*}$ & 0.159$^{*}$ & 0.167$^{*}$ & 0.164$^{*}$ & 0.232$^{*}$ & 0.226$^{*}$ &  0.235$^{*}$ \\
 &  & HR& 0.303$^{*}$ & 0.299$^{*}$ & 0.308$^{*}$ & 0.300$^{*}$ & 0.296$^{*}$ & 0.297$^{*}$ & 0.289 & 0.310$^{*}$ & 0.311$^{*}$ & 0.397$^{*}$ & 0.380$^{*}$ & 0.414$^{*}$ \\ \cmidrule[.1pt]{2-15}
 & \multirow{2}{*}{HEAD$\mathrm{^{\mathbb{H}}_P}$} & ND & \color{blue}\textbf{0.173}$^{*}$ & 0.166$^*$ & 0.169$^{*}$ & \color{blue}\textbf{0.161}$^{*}$ & 0.158$^{*}$ & 0.157$^{*}$ & 0.161$^*$ & \color{blue}\textbf{0.180}$^{*}$ & 0.175$^{*}$ &  0.238$^{*}$ & 0.232$^{*}$ &  \color{blue}\textbf{0.244}$^{*}$ \\
 &  & HR& \color{red}\textbf{0.321}$^{*}$ & 0.314$^{*}$ & 0.320$^{*}$ & \color{red}\textbf{0.309}$^{*}$ & 0.302$^{*}$ & 0.305$^{*}$ & 0.301$^{*}$ & \color{red}\textbf{0.333}$^{*}$ & 0.327$^{*}$ & 0.408$^{*}$ & 0.396$^{*}$ & \color{red}\textbf{0.417}$^{*}$ \\
\cmidrule[2pt]{1-15}
\end{tabularx}
\end{table*}

\subsection{Experimental Setup} \label{datasets}
Following the prior studies \cite{choi2022based,zhao2023crossb}, we evaluate our model using \textit{Amazon}\footnote{\url{https://cseweb.ucsd.edu/~jmcauley/datasets/amazon_v2/}} 5-core review datasets. 
As shown in Table \ref{dataset}, we take 12 domain pairs, three as the source with richer interactions and four as the target with sparse ones \cite{ben2010theory}. The pairs of \textit{(Cloth, Beauty)}, \textit{(CDs, Digital Music)}, \textit{(Toys, Video Games)} are relevant to each other. The target domain datasets are split into 80\%/10\%/10\% for training, validation, and testing without considering a temporal sequence same as \cite{choi2022based,xu2023decoupled,wang2023hdnr}. 
Additionally, we employ an early stopping technique to terminate the training process if the best validation score is not updated for 300 iterations. The dimension of word embedding is set as 100 for all methods.  

\subsection{Model Comparison (RQ1)} \label{rq1}
In Table \ref{performance_ndcg}, we show the results using Normalized Discounted Cumulative Gain (NDCG$@10$) and Hit Ratio (HR$@10$). 

\textbf{(1) Reviews and domain disentanglement enhance the quality of recommendation.} Firstly, we observe that methods utilizing only rating information \cite{he2017neural,yuan2019darec,li2020ddtcdr} show significantly lower performance compared to the review-based models. This can be due to the small size of the target domain, but we can also presume that user preferences are well reflected in the review information. We also observe that the algorithms perform differently depending on how well the reviews are utilized; attention-based AHN \cite{dong2020asymmetrical} significantly outperforms the plain text convolution model, DeepCoNN \cite{zheng2017joint}. Additionally, addressing domain discrepancies is also critical to the performance of CDR. For example, MMT \cite{krishnan2020transfer}, SER \cite{choi2022based}, and our HEAD show stable performance among the CDR techniques regardless of the source domain pairs. This can be inferred that they can separate the domain-shareable and domain-specific knowledge efficiently, making them relatively robust to noise. Additionally, our models exhibit the most stable results, suggesting that scale preservation is helpful in discriminator training.

\textbf{(2) Hyperbolic embedding achieves better performance compared to Euclidean ones, and HEAD with degree-based normalization and scale adjustment has shown its effectiveness.} 
A notable point is that hyperbolic-based methods HGCF \cite{sun2021hgcf}, GDCF \cite{zhang2022geometric}, and HDNR \cite{wang2023hdnr} exhibit good performance even without using additional domains. For example, their accuracy is comparable to AHN \cite{dong2020asymmetrical} with hierarchical attention mechanism, and DH-GAT$^{\mathbb{H}}$ \cite{xu2023decoupled} attains the best recommendation quality for some datasets. This is based on the advantages of the vast space in the hyperbolic space, which has a positive impact on learning the pairwise distance between the latent representations. In addition to this, our HEAD, which employs degree-based hierarchy correction and scale alignment, achieves a performance improvement of 10.4\% compared to SER \cite{choi2022based}. This highlights the benefits of hierarchy alignment in hyperbolic space and the removal of scale information enhances the quality of the domain discriminator.

\begin{figure}
     \centering
     \begin{subfigure}[b]{0.234\textwidth}
         \centering
         \includegraphics[width=\textwidth]{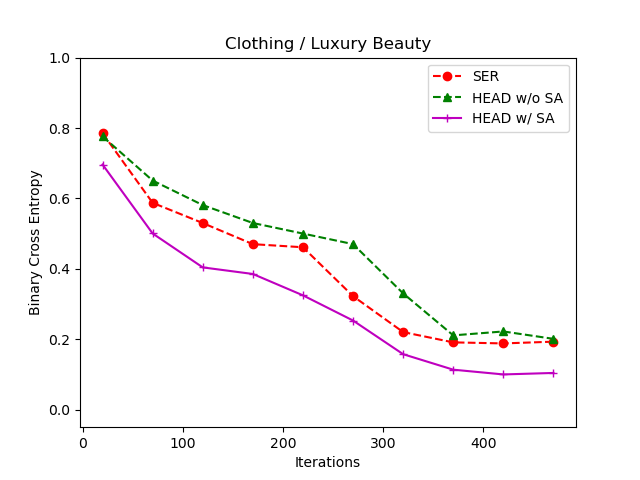}
         \caption{Similar domain}
         \label{da_loss_plot_1}
     \end{subfigure}
     \begin{subfigure}[b]{0.234\textwidth}
         \centering
         \includegraphics[width=\textwidth]{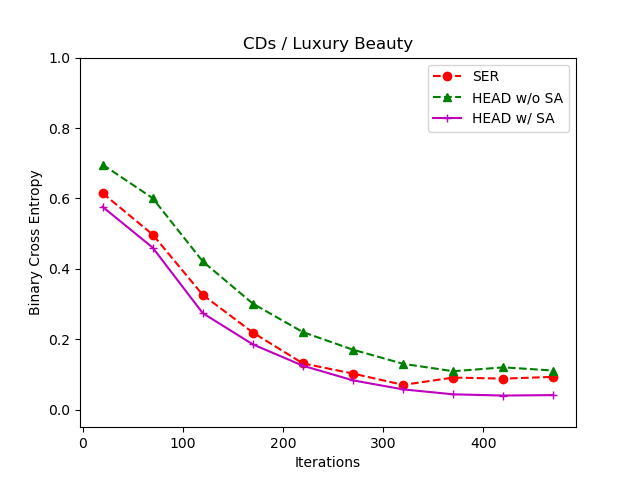}
         \caption{Dissimilar domain}
         \label{da_loss_plot_2}
     \end{subfigure}
        \caption{(RQ2) Domain discrimination performance of three methods with similar and dissimilar domain pairs}
        \label{scale_align}
\vspace{-3mm}
\end{figure}

\begin{figure}[t] 
  \includegraphics[width=.45\textwidth]{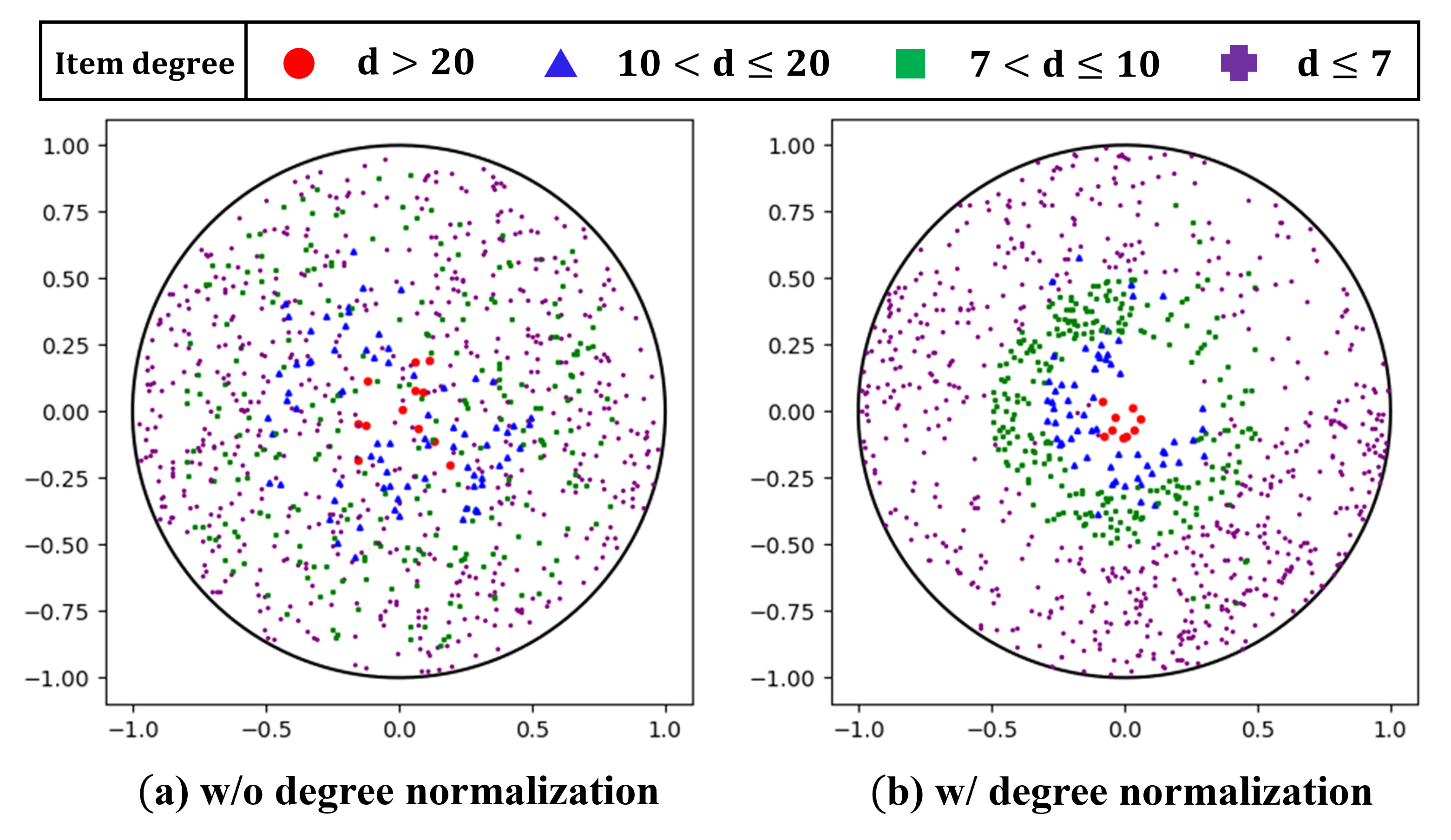}
  \caption{(RQ3) We randomly sampled 1,000 items in \textit{Digital Music} and visualized them based on their degrees}
  \label{poin_plot}
\end{figure}

\subsection{Scale Alignment and Disentanglement (RQ2)}
In Figure \ref{scale_align}, we describe the domain classification accuracy of the domain discriminator based on the application of scale alignment. Here, we employ three models: SER \cite{choi2022based} which is a state-of-the-art disentanglement algorithm, HEAD (without \textbf{S}cale \textbf{A}lignment), and HEAD (with \textbf{S}cale \textbf{A}lignment). Since the domain label is binary (0 for source and 1 for target), we describe the binary cross-entropy on the y-axis. The x-axis is the training epochs. Both figures use the \textit{Luxury Beauty} as the target domain, but each of them employs \textit{Clothing} (left) and \textit{CDs} (right) as the source domains, respectively. We discover that the discrimination accuracy is quite low in the left figure, where the two domains are similar. In addition to this, both figures represent that the discrimination accuracy of SER is better than HEAD (w/o SA), where HEAD has a larger scale than SER. This is because the hierarchical alignment in Eq. \ref{hie_emb} has a separation characteristic. However, HEAD (w/ SA) achieves the best discrimination accuracy, which confirms the proposition \ref{thm_dis}.

\subsection{Hierarchy Visualization (RQ3)}
In Figure \ref{poin_plot}, we visualize item vectors in \textit{Digital Music} dataset to assess the effect of hierarchy-aware embedding (Proposition \ref{prop_hierarchy}). Here, we randomly sample 1,000 items and classify them based on their degrees. Specifically, we average the two item vectors $S_i$ and $\widehat{S}_i$ in Eq. \ref{convolution} and project them onto the Poincaré ball (Eq. \ref{poin_ball}). The left figure employs simple root alignment (Eq. \ref{hyp_align}), while the right one further benefits from our degree-based normalization (Eq. \ref{hie_emb}). As observed in the left figure, nodes are quite randomly distributed regardless of their degrees. Although some nodes with higher degrees (red, $d > 20$) are placed near the origin, nodes with lower degrees (purple, green, and blue) are positioned quite randomly. In contrast, the right figure shows that nodes are aligned based on degrees. From this, we conclude that the degree-based normalization successfully preserves the structural information, which leads to a better utilization of hyperbolic space eventually. For case studies, we highly recommend reading this article \cite{cao2022disencdr}.

\begin{figure}
  \centering
  \includegraphics[width=0.47\textwidth]{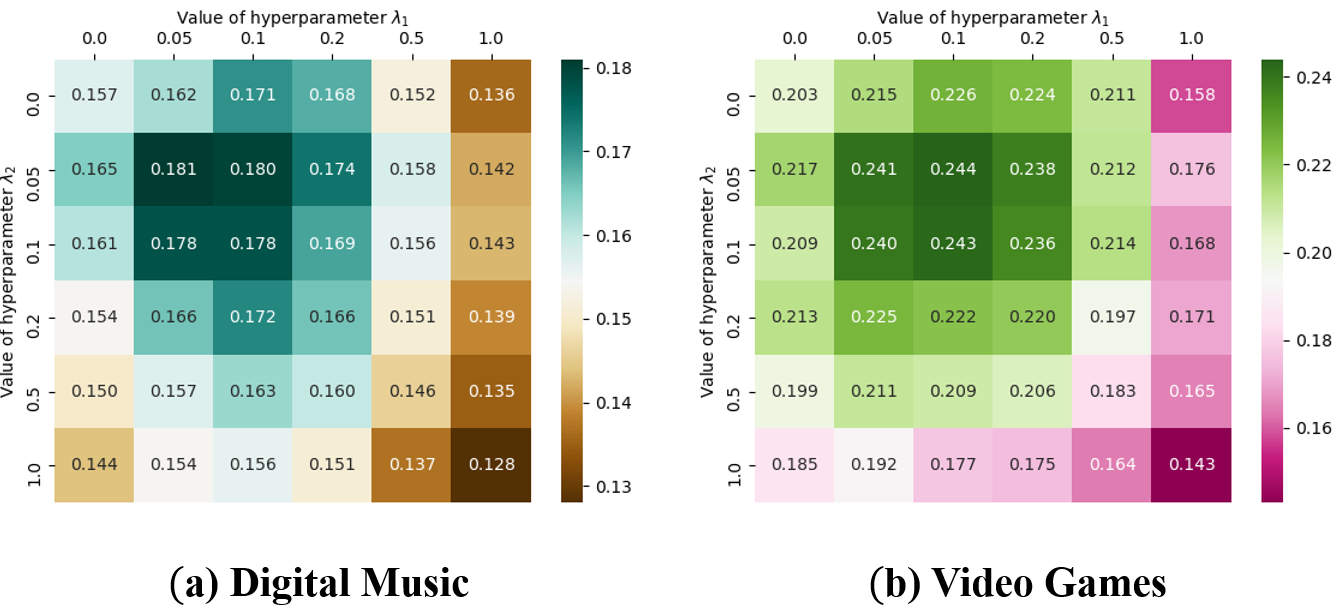}
  \caption{(RQ4) We describe the NDCG@10 of two datasets by varying the parameters $\lambda_1$ (x-axis, degree normalization) and $\lambda_2$ (y-axis, scale alignment) in Eq. \ref{total_loss}, respectively}
  \label{param_sense}
\end{figure}

\subsection{Parameter Sensitivity Analysis (RQ4)} \label{rq4}
Given the model with Poincaré Glove HEAD$^{\mathbb{H}}_{\mathbb{P}}$, we vary the weights of the loss function in Eq. \ref{total_loss}. Typically, the weight of the prediction loss (Eq. \ref{pred_loss}) is set to 1 since it is the main object of recommender systems. Now, we adjust the two hyper-parameters $\lambda_1$ and $\lambda_2$ that control the weight of hierarchy embedding and scale alignment. The experimental results are shown in Figure \ref{param_sense}, where we conduct a grid search and plot the NDCG@10 score through heatmap using the pairs of (\textit{CDs and Vinyl}, \textit{Digital Music}) and (\textit{Toys and Games}, \textit{Video Games}). Here, the rows and columns represent $\lambda_1$ and $\lambda_2$, respectively. Firstly, we observe that the performances are dismal when the hyper-parameters take large values. This is because the ranking loss is overwhelmed by other functions, making the convergence of the parameters challenging. Instead, assigning small values for both $\lambda_1$ and $\lambda_2$ enhances the overall quality of recommendation by improving structural alignment and domain disentanglement. As illustrated, we can see that setting $\lambda_1=\lambda_2=0.05$ in (a) Digital Music and $\lambda_1=0.1,\lambda_2=0.05$ in (b) Video Games achieve the best performance. One might argue that the search for optimal parameters may require huge computational costs, but we find that suppressing these values below a specific threshold grants marginal improvements only. Please refer to \cite{choi2022based} for the ablation study of using either domain-specific or shareable features.

\section{CONCLUSION}
Recent studies have addressed the challenge of data sparsity in recommender systems by integrating Cross-Domain Recommendation (CDR) with review texts. However, existing methods relying on an Euclidean space encounter difficulties due to the exponentially growing interactions between users and items. In response to this, we introduce a hyperbolic CDR as a potential solution and overcome several associated issues. Firstly, we identify some drawbacks related to root- and distance-based alignment, which are problematic in preserving the tree-like structure within a hyperbolic space. To address these issues, we propose a novel solution: hierarchy-preserving embedding and domain disentanglement. Lastly, we provide a mathematical foundation to emphasize the theoretical relevance of our proposed strategies. Experimental results demonstrate the superiority of our model over state-of-the-art single and cross-domain algorithms.

\section{Acknowledgments} 
This work was supported by Institute of Information \& Communications Technology Planning \& Evaluation (IITP) grant funded by the Korean government (MSIT) (No. 2021-0-02068, IITP-2024-00156287), and KENTECH Research Grant (202200019A) funded by the National Research Foundation of Korea (NRF) (4199990214639).


\bibliographystyle{ACM-Reference-Format}
\balance
\bibliography{references.bib}


\begin{thebibliography}{82}


\ifx \showCODEN    \undefined \def \showCODEN     #1{\unskip}     \fi
\ifx \showDOI      \undefined \def \showDOI       #1{#1}\fi
\ifx \showISBNx    \undefined \def \showISBNx     #1{\unskip}     \fi
\ifx \showISBNxiii \undefined \def \showISBNxiii  #1{\unskip}     \fi
\ifx \showISSN     \undefined \def \showISSN      #1{\unskip}     \fi
\ifx \showLCCN     \undefined \def \showLCCN      #1{\unskip}     \fi
\ifx \shownote     \undefined \def \shownote      #1{#1}          \fi
\ifx \showarticletitle \undefined \def \showarticletitle #1{#1}   \fi
\ifx \showURL      \undefined \def \showURL       {\relax}        \fi
\providecommand\bibfield[2]{#2}
\providecommand\bibinfo[2]{#2}
\providecommand\natexlab[1]{#1}
\providecommand\showeprint[2][]{arXiv:#2}

\bibitem[\protect\citeauthoryear{Achiam, Adler, Agarwal, Ahmad, Akkaya, Aleman, Almeida, Altenschmidt, Altman, Anadkat, et~al\mbox{.}}{Achiam et~al\mbox{.}}{2023}]%
        {achiam2023gpt}
\bibfield{author}{\bibinfo{person}{Josh Achiam}, \bibinfo{person}{Steven Adler}, \bibinfo{person}{Sandhini Agarwal}, \bibinfo{person}{Lama Ahmad}, \bibinfo{person}{Ilge Akkaya}, \bibinfo{person}{Florencia~Leoni Aleman}, \bibinfo{person}{Diogo Almeida}, \bibinfo{person}{Janko Altenschmidt}, \bibinfo{person}{Sam Altman}, \bibinfo{person}{Shyamal Anadkat}, {et~al\mbox{.}}} \bibinfo{year}{2023}\natexlab{}.
\newblock \showarticletitle{Gpt-4 technical report}.
\newblock \bibinfo{journal}{\emph{arXiv preprint arXiv:2303.08774}} (\bibinfo{year}{2023}).
\newblock


\bibitem[\protect\citeauthoryear{Addagarla and Amalanathan}{Addagarla and Amalanathan}{2020}]%
        {addagarla2020probabilistic}
\bibfield{author}{\bibinfo{person}{Ssvr~Kumar Addagarla} {and} \bibinfo{person}{Anthoniraj Amalanathan}.} \bibinfo{year}{2020}\natexlab{}.
\newblock \showarticletitle{Probabilistic unsupervised machine learning approach for a similar image recommender system for E-commerce}.
\newblock \bibinfo{journal}{\emph{Symmetry}} \bibinfo{volume}{12}, \bibinfo{number}{11} (\bibinfo{year}{2020}), \bibinfo{pages}{1783}.
\newblock


\bibitem[\protect\citeauthoryear{Al-Ghuribi and Noah}{Al-Ghuribi and Noah}{2019}]%
        {al2019multi}
\bibfield{author}{\bibinfo{person}{Sumaia~Mohammed Al-Ghuribi} {and} \bibinfo{person}{Shahrul Azman~Mohd Noah}.} \bibinfo{year}{2019}\natexlab{}.
\newblock \showarticletitle{Multi-criteria review-based recommender system--the state of the art}.
\newblock \bibinfo{journal}{\emph{IEEE Access}}  \bibinfo{volume}{7} (\bibinfo{year}{2019}), \bibinfo{pages}{169446--169468}.
\newblock


\bibitem[\protect\citeauthoryear{Ben-David, Blitzer, Crammer, Kulesza, Pereira, and Vaughan}{Ben-David et~al\mbox{.}}{2010}]%
        {ben2010theory}
\bibfield{author}{\bibinfo{person}{Shai Ben-David}, \bibinfo{person}{John Blitzer}, \bibinfo{person}{Koby Crammer}, \bibinfo{person}{Alex Kulesza}, \bibinfo{person}{Fernando Pereira}, {and} \bibinfo{person}{Jennifer~Wortman Vaughan}.} \bibinfo{year}{2010}\natexlab{}.
\newblock \showarticletitle{A theory of learning from different domains}.
\newblock \bibinfo{journal}{\emph{Machine learning}} \bibinfo{volume}{79}, \bibinfo{number}{1} (\bibinfo{year}{2010}), \bibinfo{pages}{151--175}.
\newblock


\bibitem[\protect\citeauthoryear{Bonab, Aliannejadi, Vardasbi, Kanoulas, and Allan}{Bonab et~al\mbox{.}}{2021}]%
        {bonab2021cross}
\bibfield{author}{\bibinfo{person}{Hamed Bonab}, \bibinfo{person}{Mohammad Aliannejadi}, \bibinfo{person}{Ali Vardasbi}, \bibinfo{person}{Evangelos Kanoulas}, {and} \bibinfo{person}{James Allan}.} \bibinfo{year}{2021}\natexlab{}.
\newblock \showarticletitle{Cross-Market Product Recommendation}. In \bibinfo{booktitle}{\emph{Proceedings of the 30th ACM International Conference on Information \& Knowledge Management}}. \bibinfo{pages}{110--119}.
\newblock


\bibitem[\protect\citeauthoryear{Bousmalis, Trigeorgis, Silberman, Krishnan, and Erhan}{Bousmalis et~al\mbox{.}}{2016}]%
        {bousmalis2016domain}
\bibfield{author}{\bibinfo{person}{Konstantinos Bousmalis}, \bibinfo{person}{George Trigeorgis}, \bibinfo{person}{Nathan Silberman}, \bibinfo{person}{Dilip Krishnan}, {and} \bibinfo{person}{Dumitru Erhan}.} \bibinfo{year}{2016}\natexlab{}.
\newblock \showarticletitle{Domain separation networks}.
\newblock \bibinfo{journal}{\emph{Advances in neural information processing systems}}  \bibinfo{volume}{29} (\bibinfo{year}{2016}), \bibinfo{pages}{343--351}.
\newblock


\bibitem[\protect\citeauthoryear{Cai, Li, Wei, Qiao, Zhang, and Hao}{Cai et~al\mbox{.}}{2019}]%
        {cai2019learning}
\bibfield{author}{\bibinfo{person}{Ruichu Cai}, \bibinfo{person}{Zijian Li}, \bibinfo{person}{Pengfei Wei}, \bibinfo{person}{Jie Qiao}, \bibinfo{person}{Kun Zhang}, {and} \bibinfo{person}{Zhifeng Hao}.} \bibinfo{year}{2019}\natexlab{}.
\newblock \showarticletitle{Learning disentangled semantic representation for domain adaptation}. In \bibinfo{booktitle}{\emph{IJCAI: proceedings of the conference}}, Vol.~\bibinfo{volume}{2019}. NIH Public Access, \bibinfo{pages}{2060}.
\newblock


\bibitem[\protect\citeauthoryear{Cao, Lin, Cong, Ya, Liu, and Wang}{Cao et~al\mbox{.}}{2022a}]%
        {cao2022disencdr}
\bibfield{author}{\bibinfo{person}{Jiangxia Cao}, \bibinfo{person}{Xixun Lin}, \bibinfo{person}{Xin Cong}, \bibinfo{person}{Jing Ya}, \bibinfo{person}{Tingwen Liu}, {and} \bibinfo{person}{Bin Wang}.} \bibinfo{year}{2022}\natexlab{a}.
\newblock \showarticletitle{Disencdr: Learning disentangled representations for cross-domain recommendation}. In \bibinfo{booktitle}{\emph{Proceedings of the 45th International ACM SIGIR Conference on Research and Development in Information Retrieval}}. \bibinfo{pages}{267--277}.
\newblock


\bibitem[\protect\citeauthoryear{Cao, Sheng, Cong, Liu, and Wang}{Cao et~al\mbox{.}}{2022b}]%
        {cao2022cross}
\bibfield{author}{\bibinfo{person}{Jiangxia Cao}, \bibinfo{person}{Jiawei Sheng}, \bibinfo{person}{Xin Cong}, \bibinfo{person}{Tingwen Liu}, {and} \bibinfo{person}{Bin Wang}.} \bibinfo{year}{2022}\natexlab{b}.
\newblock \showarticletitle{Cross-domain recommendation to cold-start users via variational information bottleneck}. In \bibinfo{booktitle}{\emph{2022 IEEE 38th International Conference on Data Engineering (ICDE)}}. IEEE, \bibinfo{pages}{2209--2223}.
\newblock


\bibitem[\protect\citeauthoryear{Chen, Zhang, Liu, and Ma}{Chen et~al\mbox{.}}{2018}]%
        {chen2018neural}
\bibfield{author}{\bibinfo{person}{Chong Chen}, \bibinfo{person}{Min Zhang}, \bibinfo{person}{Yiqun Liu}, {and} \bibinfo{person}{Shaoping Ma}.} \bibinfo{year}{2018}\natexlab{}.
\newblock \showarticletitle{Neural attentional rating regression with review-level explanations}. In \bibinfo{booktitle}{\emph{Proceedings of the 2018 World Wide Web Conference}}. \bibinfo{pages}{1583--1592}.
\newblock


\bibitem[\protect\citeauthoryear{Chen, Wang, Xu, Chen, Wang, Zhao, Li, and Wang}{Chen et~al\mbox{.}}{2023}]%
        {chen2023knowledge}
\bibfield{author}{\bibinfo{person}{Liyue Chen}, \bibinfo{person}{Linian Wang}, \bibinfo{person}{Jinyu Xu}, \bibinfo{person}{Shuai Chen}, \bibinfo{person}{Weiqiang Wang}, \bibinfo{person}{Wenbiao Zhao}, \bibinfo{person}{Qiyu Li}, {and} \bibinfo{person}{Leye Wang}.} \bibinfo{year}{2023}\natexlab{}.
\newblock \showarticletitle{Knowledge-inspired Subdomain Adaptation for Cross-Domain Knowledge Transfer}. In \bibinfo{booktitle}{\emph{Proceedings of the 32nd ACM International Conference on Information and Knowledge Management}}. \bibinfo{pages}{234--244}.
\newblock


\bibitem[\protect\citeauthoryear{Chen, Zhang, and Qin}{Chen et~al\mbox{.}}{2019b}]%
        {chen2019dynamic}
\bibfield{author}{\bibinfo{person}{Xu Chen}, \bibinfo{person}{Yongfeng Zhang}, {and} \bibinfo{person}{Zheng Qin}.} \bibinfo{year}{2019}\natexlab{b}.
\newblock \showarticletitle{Dynamic Explainable Recommendation based on Neural Attentive Models}. In \bibinfo{booktitle}{\emph{Proceedings of the AAAI Conference on Artificial Intelligence}}, Vol.~\bibinfo{volume}{33}. \bibinfo{pages}{53--60}.
\newblock


\bibitem[\protect\citeauthoryear{Chen, Vanden-Eijnden, and Bruna}{Chen et~al\mbox{.}}{2022}]%
        {chen2022feature}
\bibfield{author}{\bibinfo{person}{Zhengdao Chen}, \bibinfo{person}{Eric Vanden-Eijnden}, {and} \bibinfo{person}{Joan Bruna}.} \bibinfo{year}{2022}\natexlab{}.
\newblock \showarticletitle{On feature learning in neural networks with global convergence guarantees}.
\newblock \bibinfo{journal}{\emph{arXiv preprint arXiv:2204.10782}} (\bibinfo{year}{2022}).
\newblock


\bibitem[\protect\citeauthoryear{Chen, Wang, Xie, Wu, Bu, Wang, and Chen}{Chen et~al\mbox{.}}{2019a}]%
        {chen2019co}
\bibfield{author}{\bibinfo{person}{Zhongxia Chen}, \bibinfo{person}{Xiting Wang}, \bibinfo{person}{Xing Xie}, \bibinfo{person}{Tong Wu}, \bibinfo{person}{Guoqing Bu}, \bibinfo{person}{Yining Wang}, {and} \bibinfo{person}{Enhong Chen}.} \bibinfo{year}{2019}\natexlab{a}.
\newblock \showarticletitle{Co-attentive multi-task learning for explainable recommendation}. In \bibinfo{booktitle}{\emph{Proceedings of the 28th International Joint Conference on Artificial Intelligence}}. AAAI Press, \bibinfo{pages}{2137--2143}.
\newblock


\bibitem[\protect\citeauthoryear{Choi, Choi, Ko, Byun, and Kim}{Choi et~al\mbox{.}}{2022}]%
        {choi2022based}
\bibfield{author}{\bibinfo{person}{Yoonhyuk Choi}, \bibinfo{person}{Jiho Choi}, \bibinfo{person}{Taewook Ko}, \bibinfo{person}{Hyungho Byun}, {and} \bibinfo{person}{Chong-Kwon Kim}.} \bibinfo{year}{2022}\natexlab{}.
\newblock \showarticletitle{Based Domain Disentanglement without Duplicate Users or Contexts for Cross-Domain Recommendation}. In \bibinfo{booktitle}{\emph{Proceedings of the 31st ACM International Conference on Information \& Knowledge Management}}. \bibinfo{pages}{293--303}.
\newblock


\bibitem[\protect\citeauthoryear{Clauset, Shalizi, and Newman}{Clauset et~al\mbox{.}}{2009}]%
        {clauset2009power}
\bibfield{author}{\bibinfo{person}{Aaron Clauset}, \bibinfo{person}{Cosma~Rohilla Shalizi}, {and} \bibinfo{person}{Mark~EJ Newman}.} \bibinfo{year}{2009}\natexlab{}.
\newblock \showarticletitle{Power-law distributions in empirical data}.
\newblock \bibinfo{journal}{\emph{SIAM review}} \bibinfo{volume}{51}, \bibinfo{number}{4} (\bibinfo{year}{2009}), \bibinfo{pages}{661--703}.
\newblock


\bibitem[\protect\citeauthoryear{Dong, Ni, Cheng, Chen, Zong, Song, Liu, Chen, and De~Melo}{Dong et~al\mbox{.}}{2020}]%
        {dong2020asymmetrical}
\bibfield{author}{\bibinfo{person}{Xin Dong}, \bibinfo{person}{Jingchao Ni}, \bibinfo{person}{Wei Cheng}, \bibinfo{person}{Zhengzhang Chen}, \bibinfo{person}{Bo Zong}, \bibinfo{person}{Dongjin Song}, \bibinfo{person}{Yanchi Liu}, \bibinfo{person}{Haifeng Chen}, {and} \bibinfo{person}{Gerard De~Melo}.} \bibinfo{year}{2020}\natexlab{}.
\newblock \showarticletitle{Asymmetrical hierarchical networks with attentive interactions for interpretable review-based recommendation}. In \bibinfo{booktitle}{\emph{Proceedings of the AAAI Conference on Artificial Intelligence}}, Vol.~\bibinfo{volume}{34}. \bibinfo{pages}{7667--7674}.
\newblock


\bibitem[\protect\citeauthoryear{Elkahky, Song, and He}{Elkahky et~al\mbox{.}}{2015}]%
        {elkahky2015multi}
\bibfield{author}{\bibinfo{person}{Ali~Mamdouh Elkahky}, \bibinfo{person}{Yang Song}, {and} \bibinfo{person}{Xiaodong He}.} \bibinfo{year}{2015}\natexlab{}.
\newblock \showarticletitle{A multi-view deep learning approach for cross domain user modeling in recommendation systems}. In \bibinfo{booktitle}{\emph{Proceedings of the 24th International Conference on World Wide Web}}. \bibinfo{pages}{278--288}.
\newblock


\bibitem[\protect\citeauthoryear{Feng, Tran, Cong, Chen, Li, and Li}{Feng et~al\mbox{.}}{2020}]%
        {feng2020hme}
\bibfield{author}{\bibinfo{person}{Shanshan Feng}, \bibinfo{person}{Lucas~Vinh Tran}, \bibinfo{person}{Gao Cong}, \bibinfo{person}{Lisi Chen}, \bibinfo{person}{Jing Li}, {and} \bibinfo{person}{Fan Li}.} \bibinfo{year}{2020}\natexlab{}.
\newblock \showarticletitle{Hme: A hyperbolic metric embedding approach for next-poi recommendation}. In \bibinfo{booktitle}{\emph{Proceedings of the 43rd International ACM SIGIR Conference on research and development in information retrieval}}. \bibinfo{pages}{1429--1438}.
\newblock


\bibitem[\protect\citeauthoryear{Fu, Peng, Wang, Xu, and Li}{Fu et~al\mbox{.}}{2019}]%
        {fu2019deeply}
\bibfield{author}{\bibinfo{person}{Wenjing Fu}, \bibinfo{person}{Zhaohui Peng}, \bibinfo{person}{Senzhang Wang}, \bibinfo{person}{Yang Xu}, {and} \bibinfo{person}{Jin Li}.} \bibinfo{year}{2019}\natexlab{}.
\newblock \showarticletitle{Deeply fusing reviews and contents for cold start users in cross-domain recommendation systems}. In \bibinfo{booktitle}{\emph{Proceedings of the AAAI Conference on Artificial Intelligence}}, Vol.~\bibinfo{volume}{33}. \bibinfo{pages}{94--101}.
\newblock


\bibitem[\protect\citeauthoryear{Furusho and Ikeda}{Furusho and Ikeda}{2019}]%
        {furusho2019resnet}
\bibfield{author}{\bibinfo{person}{Yasutaka Furusho} {and} \bibinfo{person}{Kazushi Ikeda}.} \bibinfo{year}{2019}\natexlab{}.
\newblock \showarticletitle{Resnet and batch-normalization improve data separability}. In \bibinfo{booktitle}{\emph{Asian Conference on Machine Learning}}. PMLR, \bibinfo{pages}{94--108}.
\newblock


\bibitem[\protect\citeauthoryear{Globerson, Chechik, Pereira, and Tishby}{Globerson et~al\mbox{.}}{2004}]%
        {globerson2004euclidean}
\bibfield{author}{\bibinfo{person}{Amir Globerson}, \bibinfo{person}{Gal Chechik}, \bibinfo{person}{Fernando Pereira}, {and} \bibinfo{person}{Naftali Tishby}.} \bibinfo{year}{2004}\natexlab{}.
\newblock \showarticletitle{Euclidean embedding of co-occurrence data}.
\newblock \bibinfo{journal}{\emph{Advances in neural information processing systems}}  \bibinfo{volume}{17} (\bibinfo{year}{2004}).
\newblock


\bibitem[\protect\citeauthoryear{Gretton, Bousquet, Smola, and Sch{\"o}lkopf}{Gretton et~al\mbox{.}}{2005}]%
        {gretton2005measuring}
\bibfield{author}{\bibinfo{person}{Arthur Gretton}, \bibinfo{person}{Olivier Bousquet}, \bibinfo{person}{Alex Smola}, {and} \bibinfo{person}{Bernhard Sch{\"o}lkopf}.} \bibinfo{year}{2005}\natexlab{}.
\newblock \showarticletitle{Measuring statistical dependence with Hilbert-Schmidt norms}. In \bibinfo{booktitle}{\emph{International conference on algorithmic learning theory}}. Springer, \bibinfo{pages}{63--77}.
\newblock


\bibitem[\protect\citeauthoryear{Gulcehre, Denil, Malinowski, Razavi, Pascanu, Hermann, Battaglia, Bapst, Raposo, Santoro, et~al\mbox{.}}{Gulcehre et~al\mbox{.}}{2018}]%
        {gulcehre2018hyperbolic}
\bibfield{author}{\bibinfo{person}{Caglar Gulcehre}, \bibinfo{person}{Misha Denil}, \bibinfo{person}{Mateusz Malinowski}, \bibinfo{person}{Ali Razavi}, \bibinfo{person}{Razvan Pascanu}, \bibinfo{person}{Karl~Moritz Hermann}, \bibinfo{person}{Peter Battaglia}, \bibinfo{person}{Victor Bapst}, \bibinfo{person}{David Raposo}, \bibinfo{person}{Adam Santoro}, {et~al\mbox{.}}} \bibinfo{year}{2018}\natexlab{}.
\newblock \showarticletitle{Hyperbolic attention networks}.
\newblock \bibinfo{journal}{\emph{arXiv preprint arXiv:1805.09786}} (\bibinfo{year}{2018}).
\newblock


\bibitem[\protect\citeauthoryear{Guo, Tang, Chen, Zhu, Nguyen, and Yin}{Guo et~al\mbox{.}}{2021}]%
        {guo2021gcn}
\bibfield{author}{\bibinfo{person}{Lei Guo}, \bibinfo{person}{Li Tang}, \bibinfo{person}{Tong Chen}, \bibinfo{person}{Lei Zhu}, \bibinfo{person}{Quoc Viet~Hung Nguyen}, {and} \bibinfo{person}{Hongzhi Yin}.} \bibinfo{year}{2021}\natexlab{}.
\newblock \showarticletitle{DA-GCN: A Domain-aware Attentive Graph Convolution Network for Shared-account Cross-domain Sequential Recommendation}.
\newblock \bibinfo{journal}{\emph{arXiv preprint arXiv:2105.03300}} (\bibinfo{year}{2021}).
\newblock


\bibitem[\protect\citeauthoryear{Guo, Liu, Li, Ha, Ma, Wang, Zhao, Chen, and Guo}{Guo et~al\mbox{.}}{2023}]%
        {guo2023hyperbolic}
\bibfield{author}{\bibinfo{person}{Naicheng Guo}, \bibinfo{person}{Xiaolei Liu}, \bibinfo{person}{Shaoshuai Li}, \bibinfo{person}{Mingming Ha}, \bibinfo{person}{Qiongxu Ma}, \bibinfo{person}{Binfeng Wang}, \bibinfo{person}{Yunan Zhao}, \bibinfo{person}{Linxun Chen}, {and} \bibinfo{person}{Xiaobo Guo}.} \bibinfo{year}{2023}\natexlab{}.
\newblock \showarticletitle{Hyperbolic Contrastive Graph Representation Learning for Session-based Recommendation}.
\newblock \bibinfo{journal}{\emph{IEEE Transactions on Knowledge and Data Engineering}} (\bibinfo{year}{2023}).
\newblock


\bibitem[\protect\citeauthoryear{Hande, Puranik, Priyadharshini, and Chakravarthi}{Hande et~al\mbox{.}}{2021}]%
        {hande2021domain}
\bibfield{author}{\bibinfo{person}{Adeep Hande}, \bibinfo{person}{Karthik Puranik}, \bibinfo{person}{Ruba Priyadharshini}, {and} \bibinfo{person}{Bharathi~Raja Chakravarthi}.} \bibinfo{year}{2021}\natexlab{}.
\newblock \showarticletitle{Domain identification of scientific articles using transfer learning and ensembles}. In \bibinfo{booktitle}{\emph{Pacific-Asia Conference on Knowledge Discovery and Data Mining}}. Springer, \bibinfo{pages}{88--97}.
\newblock


\bibitem[\protect\citeauthoryear{He, Liao, Zhang, Nie, Hu, and Chua}{He et~al\mbox{.}}{2017}]%
        {he2017neural}
\bibfield{author}{\bibinfo{person}{Xiangnan He}, \bibinfo{person}{Lizi Liao}, \bibinfo{person}{Hanwang Zhang}, \bibinfo{person}{Liqiang Nie}, \bibinfo{person}{Xia Hu}, {and} \bibinfo{person}{Tat-Seng Chua}.} \bibinfo{year}{2017}\natexlab{}.
\newblock \showarticletitle{Neural collaborative filtering}. In \bibinfo{booktitle}{\emph{Proceedings of the 26th international conference on world wide web}}. \bibinfo{pages}{173--182}.
\newblock


\bibitem[\protect\citeauthoryear{Hu, Zhang, and Yang}{Hu et~al\mbox{.}}{2018}]%
        {hu2018conet}
\bibfield{author}{\bibinfo{person}{Guangneng Hu}, \bibinfo{person}{Yu Zhang}, {and} \bibinfo{person}{Qiang Yang}.} \bibinfo{year}{2018}\natexlab{}.
\newblock \showarticletitle{Conet: Collaborative cross networks for cross-domain recommendation}. In \bibinfo{booktitle}{\emph{Proceedings of the 27th ACM international conference on information and knowledge management}}. \bibinfo{pages}{667--676}.
\newblock


\bibitem[\protect\citeauthoryear{Kang, Hwang, Lee, and Yu}{Kang et~al\mbox{.}}{2019}]%
        {kang2019semi}
\bibfield{author}{\bibinfo{person}{SeongKu Kang}, \bibinfo{person}{Junyoung Hwang}, \bibinfo{person}{Dongha Lee}, {and} \bibinfo{person}{Hwanjo Yu}.} \bibinfo{year}{2019}\natexlab{}.
\newblock \showarticletitle{Semi-supervised learning for cross-domain recommendation to cold-start users}. In \bibinfo{booktitle}{\emph{Proceedings of the 28th ACM International Conference on Information and Knowledge Management}}. \bibinfo{pages}{1563--1572}.
\newblock


\bibitem[\protect\citeauthoryear{Kazienko, Musial, and Kajdanowicz}{Kazienko et~al\mbox{.}}{2011}]%
        {kazienko2011multidimensional}
\bibfield{author}{\bibinfo{person}{Przemys{\l}aw Kazienko}, \bibinfo{person}{Katarzyna Musial}, {and} \bibinfo{person}{Tomasz Kajdanowicz}.} \bibinfo{year}{2011}\natexlab{}.
\newblock \showarticletitle{Multidimensional social network in the social recommender system}.
\newblock \bibinfo{journal}{\emph{IEEE Transactions on Systems, Man, and Cybernetics-Part A: Systems and Humans}} \bibinfo{volume}{41}, \bibinfo{number}{4} (\bibinfo{year}{2011}), \bibinfo{pages}{746--759}.
\newblock


\bibitem[\protect\citeauthoryear{Khoshneshin and Street}{Khoshneshin and Street}{2010}]%
        {khoshneshin2010collaborative}
\bibfield{author}{\bibinfo{person}{Mohammad Khoshneshin} {and} \bibinfo{person}{W~Nick Street}.} \bibinfo{year}{2010}\natexlab{}.
\newblock \showarticletitle{Collaborative filtering via euclidean embedding}. In \bibinfo{booktitle}{\emph{Proceedings of the fourth ACM conference on Recommender systems}}. \bibinfo{pages}{87--94}.
\newblock


\bibitem[\protect\citeauthoryear{Khrulkov, Mirvakhabova, Ustinova, Oseledets, and Lempitsky}{Khrulkov et~al\mbox{.}}{2020}]%
        {khrulkov2020hyperbolic}
\bibfield{author}{\bibinfo{person}{Valentin Khrulkov}, \bibinfo{person}{Leyla Mirvakhabova}, \bibinfo{person}{Evgeniya Ustinova}, \bibinfo{person}{Ivan Oseledets}, {and} \bibinfo{person}{Victor Lempitsky}.} \bibinfo{year}{2020}\natexlab{}.
\newblock \showarticletitle{Hyperbolic image embeddings}. In \bibinfo{booktitle}{\emph{Proceedings of the IEEE/CVF Conference on Computer Vision and Pattern Recognition}}. \bibinfo{pages}{6418--6428}.
\newblock


\bibitem[\protect\citeauthoryear{Krishnan, Das, Bendre, Yang, and Sundaram}{Krishnan et~al\mbox{.}}{2020}]%
        {krishnan2020transfer}
\bibfield{author}{\bibinfo{person}{Adit Krishnan}, \bibinfo{person}{Mahashweta Das}, \bibinfo{person}{Mangesh Bendre}, \bibinfo{person}{Hao Yang}, {and} \bibinfo{person}{Hari Sundaram}.} \bibinfo{year}{2020}\natexlab{}.
\newblock \showarticletitle{Transfer Learning via Contextual Invariants for One-to-Many Cross-Domain Recommendation}. In \bibinfo{booktitle}{\emph{Proceedings of the 43rd International ACM SIGIR Conference on Research and Development in Information Retrieval}}. \bibinfo{pages}{1081--1090}.
\newblock


\bibitem[\protect\citeauthoryear{Law, Liao, Snell, and Zemel}{Law et~al\mbox{.}}{2019}]%
        {law2019lorentzian}
\bibfield{author}{\bibinfo{person}{Marc Law}, \bibinfo{person}{Renjie Liao}, \bibinfo{person}{Jake Snell}, {and} \bibinfo{person}{Richard Zemel}.} \bibinfo{year}{2019}\natexlab{}.
\newblock \showarticletitle{Lorentzian distance learning for hyperbolic representations}. In \bibinfo{booktitle}{\emph{International Conference on Machine Learning}}. PMLR, \bibinfo{pages}{3672--3681}.
\newblock


\bibitem[\protect\citeauthoryear{Li, Yang, Huo, Chen, Xu, and Wang}{Li et~al\mbox{.}}{2022}]%
        {li2022hyperbolic}
\bibfield{author}{\bibinfo{person}{Anchen Li}, \bibinfo{person}{Bo Yang}, \bibinfo{person}{Huan Huo}, \bibinfo{person}{Hongxu Chen}, \bibinfo{person}{Guandong Xu}, {and} \bibinfo{person}{Zhen Wang}.} \bibinfo{year}{2022}\natexlab{}.
\newblock \showarticletitle{Hyperbolic neural collaborative recommender}.
\newblock \bibinfo{journal}{\emph{IEEE Transactions on Knowledge and Data Engineering}} (\bibinfo{year}{2022}).
\newblock


\bibitem[\protect\citeauthoryear{Li and Tuzhilin}{Li and Tuzhilin}{2020}]%
        {li2020ddtcdr}
\bibfield{author}{\bibinfo{person}{Pan Li} {and} \bibinfo{person}{Alexander Tuzhilin}.} \bibinfo{year}{2020}\natexlab{}.
\newblock \showarticletitle{Ddtcdr: Deep dual transfer cross domain recommendation}. In \bibinfo{booktitle}{\emph{Proceedings of the 13th International Conference on Web Search and Data Mining}}. \bibinfo{pages}{331--339}.
\newblock


\bibitem[\protect\citeauthoryear{Li, Tang, Li, and He}{Li et~al\mbox{.}}{2019}]%
        {li2019learning}
\bibfield{author}{\bibinfo{person}{Zejian Li}, \bibinfo{person}{Yongchuan Tang}, \bibinfo{person}{Wei Li}, {and} \bibinfo{person}{Yongxing He}.} \bibinfo{year}{2019}\natexlab{}.
\newblock \showarticletitle{Learning disentangled representation with pairwise independence}. In \bibinfo{booktitle}{\emph{Proceedings of the AAAI Conference on Artificial Intelligence}}, Vol.~\bibinfo{volume}{33}. \bibinfo{pages}{4245--4252}.
\newblock


\bibitem[\protect\citeauthoryear{Lin and Zha}{Lin and Zha}{2008}]%
        {lin2008riemannian}
\bibfield{author}{\bibinfo{person}{Tong Lin} {and} \bibinfo{person}{Hongbin Zha}.} \bibinfo{year}{2008}\natexlab{}.
\newblock \showarticletitle{Riemannian manifold learning}.
\newblock \bibinfo{journal}{\emph{IEEE transactions on pattern analysis and machine intelligence}} \bibinfo{volume}{30}, \bibinfo{number}{5} (\bibinfo{year}{2008}), \bibinfo{pages}{796--809}.
\newblock


\bibitem[\protect\citeauthoryear{Liu, Guo, Li, Zhao, and Wu}{Liu et~al\mbox{.}}{2021}]%
        {liu2021collaborative}
\bibfield{author}{\bibinfo{person}{Huiting Liu}, \bibinfo{person}{Lingling Guo}, \bibinfo{person}{Peipei Li}, \bibinfo{person}{Peng Zhao}, {and} \bibinfo{person}{Xindong Wu}.} \bibinfo{year}{2021}\natexlab{}.
\newblock \showarticletitle{Collaborative filtering with a deep adversarial and attention network for cross-domain recommendation}.
\newblock \bibinfo{journal}{\emph{Information Sciences}}  \bibinfo{volume}{565} (\bibinfo{year}{2021}), \bibinfo{pages}{370--389}.
\newblock


\bibitem[\protect\citeauthoryear{Liu, Wang, Tang, Yang, Huang, and Liu}{Liu et~al\mbox{.}}{2019}]%
        {liu2019recommender}
\bibfield{author}{\bibinfo{person}{Tianqiao Liu}, \bibinfo{person}{Zhiwei Wang}, \bibinfo{person}{Jiliang Tang}, \bibinfo{person}{Songfan Yang}, \bibinfo{person}{Gale~Yan Huang}, {and} \bibinfo{person}{Zitao Liu}.} \bibinfo{year}{2019}\natexlab{}.
\newblock \showarticletitle{Recommender systems with heterogeneous side information}. In \bibinfo{booktitle}{\emph{The world wide web conference}}. \bibinfo{pages}{3027--3033}.
\newblock


\bibitem[\protect\citeauthoryear{Liu, Zheng, Su, Hu, Tan, and Chen}{Liu et~al\mbox{.}}{2022}]%
        {liu2022exploiting}
\bibfield{author}{\bibinfo{person}{Weiming Liu}, \bibinfo{person}{Xiaolin Zheng}, \bibinfo{person}{Jiajie Su}, \bibinfo{person}{Mengling Hu}, \bibinfo{person}{Yanchao Tan}, {and} \bibinfo{person}{Chaochao Chen}.} \bibinfo{year}{2022}\natexlab{}.
\newblock \showarticletitle{Exploiting variational domain-invariant user embedding for partially overlapped cross domain recommendation}. In \bibinfo{booktitle}{\emph{Proceedings of the 45th International ACM SIGIR Conference on Research and Development in Information Retrieval}}. \bibinfo{pages}{312--321}.
\newblock


\bibitem[\protect\citeauthoryear{Lu, Wu, Mao, Wang, and Zhang}{Lu et~al\mbox{.}}{2015}]%
        {lu2015recommender}
\bibfield{author}{\bibinfo{person}{Jie Lu}, \bibinfo{person}{Dianshuang Wu}, \bibinfo{person}{Mingsong Mao}, \bibinfo{person}{Wei Wang}, {and} \bibinfo{person}{Guangquan Zhang}.} \bibinfo{year}{2015}\natexlab{}.
\newblock \showarticletitle{Recommender system application developments: a survey}.
\newblock \bibinfo{journal}{\emph{Decision support systems}}  \bibinfo{volume}{74} (\bibinfo{year}{2015}), \bibinfo{pages}{12--32}.
\newblock


\bibitem[\protect\citeauthoryear{Man, Shen, Jin, and Cheng}{Man et~al\mbox{.}}{2017}]%
        {man2017cross}
\bibfield{author}{\bibinfo{person}{Tong Man}, \bibinfo{person}{Huawei Shen}, \bibinfo{person}{Xiaolong Jin}, {and} \bibinfo{person}{Xueqi Cheng}.} \bibinfo{year}{2017}\natexlab{}.
\newblock \showarticletitle{Cross-Domain Recommendation: An Embedding and Mapping Approach.}. In \bibinfo{booktitle}{\emph{IJCAI}}, Vol.~\bibinfo{volume}{17}. \bibinfo{pages}{2464--2470}.
\newblock


\bibitem[\protect\citeauthoryear{Mansour, Mohri, and Rostamizadeh}{Mansour et~al\mbox{.}}{2009}]%
        {mansour2009domain}
\bibfield{author}{\bibinfo{person}{Yishay Mansour}, \bibinfo{person}{Mehryar Mohri}, {and} \bibinfo{person}{Afshin Rostamizadeh}.} \bibinfo{year}{2009}\natexlab{}.
\newblock \showarticletitle{Domain adaptation: Learning bounds and algorithms}.
\newblock \bibinfo{journal}{\emph{arXiv preprint arXiv:0902.3430}} (\bibinfo{year}{2009}).
\newblock


\bibitem[\protect\citeauthoryear{Mikolov, Sutskever, Chen, Corrado, and Dean}{Mikolov et~al\mbox{.}}{2013}]%
        {mikolov2013distributed}
\bibfield{author}{\bibinfo{person}{Tomas Mikolov}, \bibinfo{person}{Ilya Sutskever}, \bibinfo{person}{Kai Chen}, \bibinfo{person}{Greg~S Corrado}, {and} \bibinfo{person}{Jeff Dean}.} \bibinfo{year}{2013}\natexlab{}.
\newblock \showarticletitle{Distributed representations of words and phrases and their compositionality}.
\newblock \bibinfo{journal}{\emph{Advances in neural information processing systems}}  \bibinfo{volume}{26} (\bibinfo{year}{2013}).
\newblock


\bibitem[\protect\citeauthoryear{Nema, Karatzoglou, and Radlinski}{Nema et~al\mbox{.}}{2021}]%
        {nema2021disentangling}
\bibfield{author}{\bibinfo{person}{Preksha Nema}, \bibinfo{person}{Alexandros Karatzoglou}, {and} \bibinfo{person}{Filip Radlinski}.} \bibinfo{year}{2021}\natexlab{}.
\newblock \showarticletitle{Disentangling Preference Representations for Recommendation Critiquing with {\ss}-VAE}. In \bibinfo{booktitle}{\emph{Proceedings of the 30th ACM International Conference on Information \& Knowledge Management}}. \bibinfo{pages}{1356--1365}.
\newblock


\bibitem[\protect\citeauthoryear{Nickel and Kiela}{Nickel and Kiela}{2017}]%
        {nickel2017poincare}
\bibfield{author}{\bibinfo{person}{Maximillian Nickel} {and} \bibinfo{person}{Douwe Kiela}.} \bibinfo{year}{2017}\natexlab{}.
\newblock \showarticletitle{Poincar{\'e} embeddings for learning hierarchical representations}.
\newblock \bibinfo{journal}{\emph{Advances in neural information processing systems}}  \bibinfo{volume}{30} (\bibinfo{year}{2017}).
\newblock


\bibitem[\protect\citeauthoryear{Nickel and Kiela}{Nickel and Kiela}{2018}]%
        {nickel2018learning}
\bibfield{author}{\bibinfo{person}{Maximillian Nickel} {and} \bibinfo{person}{Douwe Kiela}.} \bibinfo{year}{2018}\natexlab{}.
\newblock \showarticletitle{Learning continuous hierarchies in the lorentz model of hyperbolic geometry}. In \bibinfo{booktitle}{\emph{International conference on machine learning}}. PMLR, \bibinfo{pages}{3779--3788}.
\newblock


\bibitem[\protect\citeauthoryear{Peng, Huang, Sun, and Saenko}{Peng et~al\mbox{.}}{2019}]%
        {peng2019domain}
\bibfield{author}{\bibinfo{person}{Xingchao Peng}, \bibinfo{person}{Zijun Huang}, \bibinfo{person}{Ximeng Sun}, {and} \bibinfo{person}{Kate Saenko}.} \bibinfo{year}{2019}\natexlab{}.
\newblock \showarticletitle{Domain agnostic learning with disentangled representations}. In \bibinfo{booktitle}{\emph{International Conference on Machine Learning}}. PMLR, \bibinfo{pages}{5102--5112}.
\newblock


\bibitem[\protect\citeauthoryear{Pennington, Socher, and Manning}{Pennington et~al\mbox{.}}{2014}]%
        {pennington2014glove}
\bibfield{author}{\bibinfo{person}{Jeffrey Pennington}, \bibinfo{person}{Richard Socher}, {and} \bibinfo{person}{Christopher~D Manning}.} \bibinfo{year}{2014}\natexlab{}.
\newblock \showarticletitle{Glove: Global vectors for word representation}. In \bibinfo{booktitle}{\emph{Proceedings of the 2014 conference on empirical methods in natural language processing (EMNLP)}}. \bibinfo{pages}{1532--1543}.
\newblock


\bibitem[\protect\citeauthoryear{Ramakrishnan, Agrawal, and Lee}{Ramakrishnan et~al\mbox{.}}{2018}]%
        {ramakrishnan2018overcoming}
\bibfield{author}{\bibinfo{person}{Sainandan Ramakrishnan}, \bibinfo{person}{Aishwarya Agrawal}, {and} \bibinfo{person}{Stefan Lee}.} \bibinfo{year}{2018}\natexlab{}.
\newblock \showarticletitle{Overcoming language priors in visual question answering with adversarial regularization}.
\newblock \bibinfo{journal}{\emph{arXiv preprint arXiv:1810.03649}} (\bibinfo{year}{2018}).
\newblock


\bibitem[\protect\citeauthoryear{Sachdeva and McAuley}{Sachdeva and McAuley}{2020}]%
        {sachdeva2020useful}
\bibfield{author}{\bibinfo{person}{Noveen Sachdeva} {and} \bibinfo{person}{Julian McAuley}.} \bibinfo{year}{2020}\natexlab{}.
\newblock \showarticletitle{How Useful are Reviews for Recommendation? A Critical Review and Potential Improvements}. In \bibinfo{booktitle}{\emph{Proceedings of the 43rd International ACM SIGIR Conference on Research and Development in Information Retrieval}}. \bibinfo{pages}{1845--1848}.
\newblock


\bibitem[\protect\citeauthoryear{Seo, Huang, Yang, and Liu}{Seo et~al\mbox{.}}{2017}]%
        {seo2017interpretable}
\bibfield{author}{\bibinfo{person}{Sungyong Seo}, \bibinfo{person}{Jing Huang}, \bibinfo{person}{Hao Yang}, {and} \bibinfo{person}{Yan Liu}.} \bibinfo{year}{2017}\natexlab{}.
\newblock \showarticletitle{Interpretable convolutional neural networks with dual local and global attention for review rating prediction}. In \bibinfo{booktitle}{\emph{Proceedings of the eleventh ACM conference on recommender systems}}. \bibinfo{pages}{297--305}.
\newblock


\bibitem[\protect\citeauthoryear{Srifi, Oussous, Ait~Lahcen, and Mouline}{Srifi et~al\mbox{.}}{2020}]%
        {srifi2020recommender}
\bibfield{author}{\bibinfo{person}{Mehdi Srifi}, \bibinfo{person}{Ahmed Oussous}, \bibinfo{person}{Ayoub Ait~Lahcen}, {and} \bibinfo{person}{Salma Mouline}.} \bibinfo{year}{2020}\natexlab{}.
\newblock \showarticletitle{Recommender systems based on collaborative filtering using review texts—a survey}.
\newblock \bibinfo{journal}{\emph{Information}} \bibinfo{volume}{11}, \bibinfo{number}{6} (\bibinfo{year}{2020}), \bibinfo{pages}{317}.
\newblock


\bibitem[\protect\citeauthoryear{Su, Chen, Liu, Wu, Zheng, and Lyu}{Su et~al\mbox{.}}{2023}]%
        {su2023enhancing}
\bibfield{author}{\bibinfo{person}{Jiajie Su}, \bibinfo{person}{Chaochao Chen}, \bibinfo{person}{Weiming Liu}, \bibinfo{person}{Fei Wu}, \bibinfo{person}{Xiaolin Zheng}, {and} \bibinfo{person}{Haoming Lyu}.} \bibinfo{year}{2023}\natexlab{}.
\newblock \showarticletitle{Enhancing Hierarchy-Aware Graph Networks with Deep Dual Clustering for Session-based Recommendation}. In \bibinfo{booktitle}{\emph{Proceedings of the ACM Web Conference 2023}}. \bibinfo{pages}{165--176}.
\newblock


\bibitem[\protect\citeauthoryear{Sun, Cheng, Zuberi, P{\'e}rez, and Volkovs}{Sun et~al\mbox{.}}{2021}]%
        {sun2021hgcf}
\bibfield{author}{\bibinfo{person}{Jianing Sun}, \bibinfo{person}{Zhaoyue Cheng}, \bibinfo{person}{Saba Zuberi}, \bibinfo{person}{Felipe P{\'e}rez}, {and} \bibinfo{person}{Maksims Volkovs}.} \bibinfo{year}{2021}\natexlab{}.
\newblock \showarticletitle{Hgcf: Hyperbolic graph convolution networks for collaborative filtering}. In \bibinfo{booktitle}{\emph{Proceedings of the Web Conference 2021}}. \bibinfo{pages}{593--601}.
\newblock


\bibitem[\protect\citeauthoryear{Tay, Luu, and Hui}{Tay et~al\mbox{.}}{2018}]%
        {tay2018multi}
\bibfield{author}{\bibinfo{person}{Yi Tay}, \bibinfo{person}{Anh~Tuan Luu}, {and} \bibinfo{person}{Siu~Cheung Hui}.} \bibinfo{year}{2018}\natexlab{}.
\newblock \showarticletitle{Multi-pointer co-attention networks for recommendation}. In \bibinfo{booktitle}{\emph{Proceedings of the 24th ACM SIGKDD International Conference on Knowledge Discovery \& Data Mining}}. \bibinfo{pages}{2309--2318}.
\newblock


\bibitem[\protect\citeauthoryear{Tifrea, B{\'e}cigneul, and Ganea}{Tifrea et~al\mbox{.}}{2018}]%
        {tifrea2018poincar}
\bibfield{author}{\bibinfo{person}{Alexandru Tifrea}, \bibinfo{person}{Gary B{\'e}cigneul}, {and} \bibinfo{person}{Octavian-Eugen Ganea}.} \bibinfo{year}{2018}\natexlab{}.
\newblock \showarticletitle{Poincar$\backslash$'e glove: Hyperbolic word embeddings}.
\newblock \bibinfo{journal}{\emph{arXiv preprint arXiv:1810.06546}} (\bibinfo{year}{2018}).
\newblock


\bibitem[\protect\citeauthoryear{Touvron, Lavril, Izacard, Martinet, Lachaux, Lacroix, Rozi{\`e}re, Goyal, Hambro, Azhar, et~al\mbox{.}}{Touvron et~al\mbox{.}}{2023}]%
        {touvron2023llama}
\bibfield{author}{\bibinfo{person}{Hugo Touvron}, \bibinfo{person}{Thibaut Lavril}, \bibinfo{person}{Gautier Izacard}, \bibinfo{person}{Xavier Martinet}, \bibinfo{person}{Marie-Anne Lachaux}, \bibinfo{person}{Timoth{\'e}e Lacroix}, \bibinfo{person}{Baptiste Rozi{\`e}re}, \bibinfo{person}{Naman Goyal}, \bibinfo{person}{Eric Hambro}, \bibinfo{person}{Faisal Azhar}, {et~al\mbox{.}}} \bibinfo{year}{2023}\natexlab{}.
\newblock \showarticletitle{Llama: Open and efficient foundation language models}.
\newblock \bibinfo{journal}{\emph{arXiv preprint arXiv:2302.13971}} (\bibinfo{year}{2023}).
\newblock


\bibitem[\protect\citeauthoryear{Vinh~Tran, Tay, Zhang, Cong, and Li}{Vinh~Tran et~al\mbox{.}}{2020}]%
        {vinh2020hyperml}
\bibfield{author}{\bibinfo{person}{Lucas Vinh~Tran}, \bibinfo{person}{Yi Tay}, \bibinfo{person}{Shuai Zhang}, \bibinfo{person}{Gao Cong}, {and} \bibinfo{person}{Xiaoli Li}.} \bibinfo{year}{2020}\natexlab{}.
\newblock \showarticletitle{Hyperml: A boosting metric learning approach in hyperbolic space for recommender systems}. In \bibinfo{booktitle}{\emph{Proceedings of the 13th international conference on web search and data mining}}. \bibinfo{pages}{609--617}.
\newblock


\bibitem[\protect\citeauthoryear{Wan}{Wan}{2019}]%
        {wan2019influence}
\bibfield{author}{\bibinfo{person}{Xing Wan}.} \bibinfo{year}{2019}\natexlab{}.
\newblock \showarticletitle{Influence of feature scaling on convergence of gradient iterative algorithm}. In \bibinfo{booktitle}{\emph{Journal of physics: Conference series}}, Vol.~\bibinfo{volume}{1213}. IOP Publishing, \bibinfo{pages}{032021}.
\newblock


\bibitem[\protect\citeauthoryear{Wang, Lian, Tong, Liu, Huang, and Chen}{Wang et~al\mbox{.}}{2021}]%
        {wang2021hypersorec}
\bibfield{author}{\bibinfo{person}{Hao Wang}, \bibinfo{person}{Defu Lian}, \bibinfo{person}{Hanghang Tong}, \bibinfo{person}{Qi Liu}, \bibinfo{person}{Zhenya Huang}, {and} \bibinfo{person}{Enhong Chen}.} \bibinfo{year}{2021}\natexlab{}.
\newblock \showarticletitle{Hypersorec: Exploiting hyperbolic user and item representations with multiple aspects for social-aware recommendation}.
\newblock \bibinfo{journal}{\emph{ACM Transactions on Information Systems (TOIS)}} \bibinfo{volume}{40}, \bibinfo{number}{2} (\bibinfo{year}{2021}), \bibinfo{pages}{1--28}.
\newblock


\bibitem[\protect\citeauthoryear{Wang, Guo, Wang, Liu, and Xu}{Wang et~al\mbox{.}}{2023}]%
        {wang2023hdnr}
\bibfield{author}{\bibinfo{person}{Shicheng Wang}, \bibinfo{person}{Shu Guo}, \bibinfo{person}{Lihong Wang}, \bibinfo{person}{Tingwen Liu}, {and} \bibinfo{person}{Hongbo Xu}.} \bibinfo{year}{2023}\natexlab{}.
\newblock \showarticletitle{HDNR: A Hyperbolic-Based Debiased Approach for Personalized News Recommendation}. In \bibinfo{booktitle}{\emph{Proceedings of the 46th International ACM SIGIR Conference on Research and Development in Information Retrieval}}. \bibinfo{pages}{259--268}.
\newblock


\bibitem[\protect\citeauthoryear{Wang, Peng, Wang, Philip, Fu, and Hong}{Wang et~al\mbox{.}}{2018}]%
        {wang2018cross}
\bibfield{author}{\bibinfo{person}{Xinghua Wang}, \bibinfo{person}{Zhaohui Peng}, \bibinfo{person}{Senzhang Wang}, \bibinfo{person}{S~Yu Philip}, \bibinfo{person}{Wenjing Fu}, {and} \bibinfo{person}{Xiaoguang Hong}.} \bibinfo{year}{2018}\natexlab{}.
\newblock \showarticletitle{Cross-domain recommendation for cold-start users via neighborhood based feature mapping}. In \bibinfo{booktitle}{\emph{International conference on database systems for advanced applications}}. Springer, \bibinfo{pages}{158--165}.
\newblock


\bibitem[\protect\citeauthoryear{Xie, Liu, Wang, Liu, Zhang, and Lin}{Xie et~al\mbox{.}}{2022}]%
        {xie2022contrastive}
\bibfield{author}{\bibinfo{person}{Ruobing Xie}, \bibinfo{person}{Qi Liu}, \bibinfo{person}{Liangdong Wang}, \bibinfo{person}{Shukai Liu}, \bibinfo{person}{Bo Zhang}, {and} \bibinfo{person}{Leyu Lin}.} \bibinfo{year}{2022}\natexlab{}.
\newblock \showarticletitle{Contrastive cross-domain recommendation in matching}. In \bibinfo{booktitle}{\emph{Proceedings of the 28th ACM SIGKDD Conference on Knowledge Discovery and Data Mining}}. \bibinfo{pages}{4226--4236}.
\newblock


\bibitem[\protect\citeauthoryear{Xu and Cai}{Xu and Cai}{2023}]%
        {xu2023decoupled}
\bibfield{author}{\bibinfo{person}{Jingyun Xu} {and} \bibinfo{person}{Yi Cai}.} \bibinfo{year}{2023}\natexlab{}.
\newblock \showarticletitle{Decoupled Hyperbolic Graph Attention Network for Cross-domain Named Entity Recognition}. In \bibinfo{booktitle}{\emph{Proceedings of the 46th International ACM SIGIR Conference on Research and Development in Information Retrieval}}. \bibinfo{pages}{591--600}.
\newblock


\bibitem[\protect\citeauthoryear{Yan, Hashemi, Swersky, Yang, and Koutra}{Yan et~al\mbox{.}}{2022}]%
        {yan2022two}
\bibfield{author}{\bibinfo{person}{Yujun Yan}, \bibinfo{person}{Milad Hashemi}, \bibinfo{person}{Kevin Swersky}, \bibinfo{person}{Yaoqing Yang}, {and} \bibinfo{person}{Danai Koutra}.} \bibinfo{year}{2022}\natexlab{}.
\newblock \showarticletitle{Two sides of the same coin: Heterophily and oversmoothing in graph convolutional neural networks}. In \bibinfo{booktitle}{\emph{2022 IEEE International Conference on Data Mining (ICDM)}}. IEEE, \bibinfo{pages}{1287--1292}.
\newblock


\bibitem[\protect\citeauthoryear{Yang, Li, Zhou, Liu, and King}{Yang et~al\mbox{.}}{2022a}]%
        {yang2022hicf}
\bibfield{author}{\bibinfo{person}{Menglin Yang}, \bibinfo{person}{Zhihao Li}, \bibinfo{person}{Min Zhou}, \bibinfo{person}{Jiahong Liu}, {and} \bibinfo{person}{Irwin King}.} \bibinfo{year}{2022}\natexlab{a}.
\newblock \showarticletitle{Hicf: Hyperbolic informative collaborative filtering}. In \bibinfo{booktitle}{\emph{Proceedings of the 28th ACM SIGKDD Conference on Knowledge Discovery and Data Mining}}. \bibinfo{pages}{2212--2221}.
\newblock


\bibitem[\protect\citeauthoryear{Yang, Zhou, Liu, Lian, and King}{Yang et~al\mbox{.}}{2022b}]%
        {yang2022hrcf}
\bibfield{author}{\bibinfo{person}{Menglin Yang}, \bibinfo{person}{Min Zhou}, \bibinfo{person}{Jiahong Liu}, \bibinfo{person}{Defu Lian}, {and} \bibinfo{person}{Irwin King}.} \bibinfo{year}{2022}\natexlab{b}.
\newblock \showarticletitle{HRCF: Enhancing collaborative filtering via hyperbolic geometric regularization}. In \bibinfo{booktitle}{\emph{Proceedings of the ACM Web Conference 2022}}. \bibinfo{pages}{2462--2471}.
\newblock


\bibitem[\protect\citeauthoryear{Yang, Liu, Su, Tang, Liu, and He}{Yang et~al\mbox{.}}{2021}]%
        {yang2021autoft}
\bibfield{author}{\bibinfo{person}{Xiangli Yang}, \bibinfo{person}{Qing Liu}, \bibinfo{person}{Rong Su}, \bibinfo{person}{Ruiming Tang}, \bibinfo{person}{Zhirong Liu}, {and} \bibinfo{person}{Xiuqiang He}.} \bibinfo{year}{2021}\natexlab{}.
\newblock \showarticletitle{AutoFT: Automatic Fine-Tune for Parameters Transfer Learning in Click-Through Rate Prediction}.
\newblock \bibinfo{journal}{\emph{arXiv preprint arXiv:2106.04873}} (\bibinfo{year}{2021}).
\newblock


\bibitem[\protect\citeauthoryear{Yuan, He, Karatzoglou, and Zhang}{Yuan et~al\mbox{.}}{2020}]%
        {yuan2020parameter}
\bibfield{author}{\bibinfo{person}{Fajie Yuan}, \bibinfo{person}{Xiangnan He}, \bibinfo{person}{Alexandros Karatzoglou}, {and} \bibinfo{person}{Liguang Zhang}.} \bibinfo{year}{2020}\natexlab{}.
\newblock \showarticletitle{Parameter-efficient transfer from sequential behaviors for user modeling and recommendation}. In \bibinfo{booktitle}{\emph{Proceedings of the 43rd International ACM SIGIR Conference on Research and Development in Information Retrieval}}. \bibinfo{pages}{1469--1478}.
\newblock


\bibitem[\protect\citeauthoryear{Yuan, Yao, and Benatallah}{Yuan et~al\mbox{.}}{2019}]%
        {yuan2019darec}
\bibfield{author}{\bibinfo{person}{Feng Yuan}, \bibinfo{person}{Lina Yao}, {and} \bibinfo{person}{Boualem Benatallah}.} \bibinfo{year}{2019}\natexlab{}.
\newblock \showarticletitle{DARec: deep domain adaptation for cross-domain recommendation via transferring rating patterns}.
\newblock \bibinfo{journal}{\emph{arXiv preprint arXiv:1905.10760}} (\bibinfo{year}{2019}).
\newblock


\bibitem[\protect\citeauthoryear{Zeng, Xu, and Ai}{Zeng et~al\mbox{.}}{2021}]%
        {zeng2021zero}
\bibfield{author}{\bibinfo{person}{Hansi Zeng}, \bibinfo{person}{Zhichao Xu}, {and} \bibinfo{person}{Qingyao Ai}.} \bibinfo{year}{2021}\natexlab{}.
\newblock \showarticletitle{A Zero Attentive Relevance Matching Networkfor Review Modeling in Recommendation System}.
\newblock \bibinfo{journal}{\emph{arXiv preprint arXiv:2101.06387}} (\bibinfo{year}{2021}).
\newblock


\bibitem[\protect\citeauthoryear{Zhang, Yao, Sun, and Tay}{Zhang et~al\mbox{.}}{2019}]%
        {zhang2019deep}
\bibfield{author}{\bibinfo{person}{Shuai Zhang}, \bibinfo{person}{Lina Yao}, \bibinfo{person}{Aixin Sun}, {and} \bibinfo{person}{Yi Tay}.} \bibinfo{year}{2019}\natexlab{}.
\newblock \showarticletitle{Deep learning based recommender system: A survey and new perspectives}.
\newblock \bibinfo{journal}{\emph{ACM computing surveys (CSUR)}} \bibinfo{volume}{52}, \bibinfo{number}{1} (\bibinfo{year}{2019}), \bibinfo{pages}{1--38}.
\newblock


\bibitem[\protect\citeauthoryear{Zhang, Li, Xie, Wang, Shi, Liu, Sun, Zhang, Deng, and Zhang}{Zhang et~al\mbox{.}}{2022}]%
        {zhang2022geometric}
\bibfield{author}{\bibinfo{person}{Yiding Zhang}, \bibinfo{person}{Chaozhuo Li}, \bibinfo{person}{Xing Xie}, \bibinfo{person}{Xiao Wang}, \bibinfo{person}{Chuan Shi}, \bibinfo{person}{Yuming Liu}, \bibinfo{person}{Hao Sun}, \bibinfo{person}{Liangjie Zhang}, \bibinfo{person}{Weiwei Deng}, {and} \bibinfo{person}{Qi Zhang}.} \bibinfo{year}{2022}\natexlab{}.
\newblock \showarticletitle{Geometric disentangled collaborative filtering}. In \bibinfo{booktitle}{\emph{Proceedings of the 45th International ACM SIGIR Conference on Research and Development in Information Retrieval}}. \bibinfo{pages}{80--90}.
\newblock


\bibitem[\protect\citeauthoryear{Zhao, Li, Xiao, Deng, and Sun}{Zhao et~al\mbox{.}}{2020}]%
        {zhao2020catn}
\bibfield{author}{\bibinfo{person}{Cheng Zhao}, \bibinfo{person}{Chenliang Li}, \bibinfo{person}{Rong Xiao}, \bibinfo{person}{Hongbo Deng}, {and} \bibinfo{person}{Aixin Sun}.} \bibinfo{year}{2020}\natexlab{}.
\newblock \showarticletitle{CATN: Cross-Domain Recommendation for Cold-Start Users via Aspect Transfer Network}.
\newblock \bibinfo{journal}{\emph{arXiv preprint arXiv:2005.10549}} (\bibinfo{year}{2020}).
\newblock


\bibitem[\protect\citeauthoryear{Zhao, Zhao, He, Zhang, and Fan}{Zhao et~al\mbox{.}}{2023a}]%
        {zhao2023crossb}
\bibfield{author}{\bibinfo{person}{Chuang Zhao}, \bibinfo{person}{Hongke Zhao}, \bibinfo{person}{Ming He}, \bibinfo{person}{Jian Zhang}, {and} \bibinfo{person}{Jianping Fan}.} \bibinfo{year}{2023}\natexlab{a}.
\newblock \showarticletitle{Cross-domain recommendation via user interest alignment}. In \bibinfo{booktitle}{\emph{Proceedings of the ACM Web Conference 2023}}. \bibinfo{pages}{887--896}.
\newblock


\bibitem[\protect\citeauthoryear{Zhao, Zhao, Li, He, Wang, and Fan}{Zhao et~al\mbox{.}}{2023b}]%
        {zhao2023crossa}
\bibfield{author}{\bibinfo{person}{Chuang Zhao}, \bibinfo{person}{Hongke Zhao}, \bibinfo{person}{Xiaomeng Li}, \bibinfo{person}{Ming He}, \bibinfo{person}{Jiahui Wang}, {and} \bibinfo{person}{Jianping Fan}.} \bibinfo{year}{2023}\natexlab{b}.
\newblock \showarticletitle{Cross-domain recommendation via progressive structural alignment}.
\newblock \bibinfo{journal}{\emph{IEEE Transactions on Knowledge and Data Engineering}} (\bibinfo{year}{2023}).
\newblock


\bibitem[\protect\citeauthoryear{Zheng, Noroozi, and Yu}{Zheng et~al\mbox{.}}{2017}]%
        {zheng2017joint}
\bibfield{author}{\bibinfo{person}{Lei Zheng}, \bibinfo{person}{Vahid Noroozi}, {and} \bibinfo{person}{Philip~S Yu}.} \bibinfo{year}{2017}\natexlab{}.
\newblock \showarticletitle{Joint deep modeling of users and items using reviews for recommendation}. In \bibinfo{booktitle}{\emph{Proceedings of the Tenth ACM International Conference on Web Search and Data Mining}}. \bibinfo{pages}{425--434}.
\newblock


\bibitem[\protect\citeauthoryear{Zhu, Wang, Chen, Zhou, Li, and Liu}{Zhu et~al\mbox{.}}{2021}]%
        {zhu2021cross}
\bibfield{author}{\bibinfo{person}{Feng Zhu}, \bibinfo{person}{Yan Wang}, \bibinfo{person}{Chaochao Chen}, \bibinfo{person}{Jun Zhou}, \bibinfo{person}{Longfei Li}, {and} \bibinfo{person}{Guanfeng Liu}.} \bibinfo{year}{2021}\natexlab{}.
\newblock \showarticletitle{Cross-domain recommendation: challenges, progress, and prospects}.
\newblock \bibinfo{journal}{\emph{arXiv preprint arXiv:2103.01696}} (\bibinfo{year}{2021}).
\newblock


\bibitem[\protect\citeauthoryear{Zhu, Tang, Liu, Zhuang, Xie, Zhang, Lin, and He}{Zhu et~al\mbox{.}}{2022}]%
        {zhu2022personalized}
\bibfield{author}{\bibinfo{person}{Yongchun Zhu}, \bibinfo{person}{Zhenwei Tang}, \bibinfo{person}{Yudan Liu}, \bibinfo{person}{Fuzhen Zhuang}, \bibinfo{person}{Ruobing Xie}, \bibinfo{person}{Xu Zhang}, \bibinfo{person}{Leyu Lin}, {and} \bibinfo{person}{Qing He}.} \bibinfo{year}{2022}\natexlab{}.
\newblock \showarticletitle{Personalized transfer of user preferences for cross-domain recommendation}. In \bibinfo{booktitle}{\emph{Proceedings of the Fifteenth ACM International Conference on Web Search and Data Mining}}. \bibinfo{pages}{1507--1515}.
\newblock


\end{thebibliography}

\end{document}